\documentstyle[graphicx, amssymb, amsmath, psfig, epsfig]{mn}

\title[3D weak shear systematics]{Systematic effects on dark energy from 3D weak shear}
\author[T. D. Kitching et al.]
       {T. D. Kitching\thanks{tdk@astro.ox.ac.uk}$^{1}$,
	A. N. Taylor$^{2}$, A. F. Heavens$^{2}$
\\
$^{1}$University of Oxford, Denys Wilkinson Building, Department of Physics, 
Wilkinson Building, Keble Road, Oxford OX1 3RH, U.K.\\
$^{2}$SUPA\thanks{The Scottish Universities Physics Alliance}, Institute
for Astronomy, University of Edinburgh, Royal Observatory, Blackford
Hill, Edinburgh, EH9 3HJ, U.K.\\}

\newcommand{\be}{\begin{equation}}
\newcommand{\ee}{\end{equation}}
\newcommand{\ba}{\begin{eqnarray}}
\newcommand{\ea}{\end{eqnarray}}
\newcommand{\nn}{\nonumber \\}

\newcommand{\lgl}{\langle}
\newcommand{\rgl}{\rangle}

\newcommand{\Phib}{\mbox{\boldmath $\Phi$}}

\newcommand{\F}{\mbox{\boldmath $F$}}

\newcommand{\bell}{{\mbox{\boldmath{$\ell$}}}}

\def\gs{\mathrel{\raise1.16pt\hbox{$>$}\kern-7.0pt %
\lower3.06pt\hbox{{$\scriptstyle \sim$}}}}         %
\def\ls{\mathrel{\raise1.16pt\hbox{$<$}\kern-7.0pt %
\lower3.06pt\hbox{{$\scriptstyle \sim$}}}}         %

\begin{document}

\maketitle

\begin{abstract}
We present an investigation into the potential effect of systematics inherent in multi-band 
wide field surveys on the dark energy equation 
of state determination for two 3D weak lensing methods. The weak lensing methods are a 
geometric shear-ratio method and 3D cosmic shear. 
The analysis here uses an extension of the Fisher matrix framework 
to jointly include photometric redshift systematics, shear distortion 
systematics and 
intrinsic alignments. Using analytic parameterisations of these three primary systematic effects 
allows an isolation of systematic parameters of particular importance. 

We show that assuming systematic parameters are fixed, but possibly biased, 
results in potentially large biases in dark energy parameters. 
We quantify any potential bias by defining a Bias Figure of Merit. By marginalising over 
extra systematic parameters such biases are negated at the expense of an increase in the 
cosmological parameter errors. We show the effect on the dark energy Figure of Merit 
of marginalising over each systematic parameter 
individually. We also show the overall reduction in the 
Figure of Merit due to all three types of systematic effects. 

Based on some assumption of the likely level of systematic errors, 
we find that the largest effect on the Figure of Merit comes from 
uncertainty in the photometric redshift systematic parameters.  
These can reduce the Figure of Merit by up to a factor of $2$ to $4$ in 
both 3D weak lensing methods, if no informative prior on the systematic parameters 
is applied. Shear distortion systematics have a smaller overall effect. 
Intrinsic alignment effects can reduce the Figure of Merit by up to a further factor of $2$. 
This, however, is a worst case scenario. By including prior information 
on systematic parameters the Figure of Merit 
can be recovered to a large extent, and combined constraints from 3D cosmic shear and shear ratio 
are robust to systematics. We conclude that, as a rule of thumb, 
given a realistic current understanding of 
intrinsic alignments and photometric redshifts, then including all three primary 
systematic effects reduces the Figure of Merit by \emph{at most} a factor of $2$. 
\end{abstract}


\section{Introduction}
\label{Introduction}
It has recently been shown that the equation of state of dark energy could be 
constrained to a high degree of accuracy using wide and deep imaging surveys (see 
Albrecht et al., 2006; Peacock et al., 2006; for recent and extensive reviews).
3D weak lensing has been shown to be a particularly powerful way to use the information 
from such surveys in the determination of dark energy parameters 
(see Munshi et al., 2006 for a recent review). 3D 
weak lensing, in which the shear and redshift information of every galaxy is used, has the potential 
to constrain the dark energy equation of state, $w(z)=\rho_{\rm de}(z)/p_{\rm de}(z)$, to 
$\Delta w(z)\sim 0.01$ using surveys such as Pan-STARRS
(Kaiser et al., 2002) and DUNE (Refregier et al., 2006). 
However, the predictions made thus far (Heavens et al., 2006; Taylor et al., 2007) 
have only included statistical errors and have not included systematic effects. 
Since the scientific goal of many future surveys is to constrain $\Delta w(z)\sim 0.01$ 
such systematic effects have the potential to render any cosmological constraints impotent. 

In this paper we will address astrophysical, instrumental and theoretical systematic 
effects relevant to multi-band weak lensing surveys in an analytic way.
We specifically study the systematic effects of photometric redshifts, intrinsic alignments and  
shear distortion. We consider these three primary systematics to 
have potentially the largest effect on the ability 
of weak lensing surveys to constrain cosmological parameters. Secondary effects, such as 
source clustering, non-Gaussian effects and theoretical approximations such as the 
Born approximation have been shown 
to have a smaller effect on shear statistics (Shapiro \& Cooray, 2006; Semboloni et al., 2007; 
Schneider et al., 2002). Note that 
the effect of  non-Gaussianity (Semboloni et al., 2007) has only been 
studied via simulations, a full analytic investigation could reveal non-Gaussianity 
to be a larger systematic effect (Takada, private communication).  
We will examine the degradation that these primary effects may produce in the 
determination of the dark energy equation of state constraints for two 3D weak lensing methods: 
3D cosmic shear (Heavens, 2003; Castro et al., 2005; Heavens et al., 2006) and the shear-ratio 
method (Jain \& Taylor, 2003; Taylor et al., 2007). 

The spirit of the approach taken here is to use simple, analytic, descriptions of systematic 
effects. By distilling complex effects into simple components the change in cosmological
parameter determination due to any 
particular aspect of a systematic effect can be analysed independently. For example, is the 
bias or the fraction of outliers in photometric redshifts 
a more important factor? The obvious penalty in taking such an 
approach is that the analytic approximations made may not be fully representative of real 
systematic effects. 

The two 3D weak lensing methods are introduced 
in Section \ref{Methodology}, however we urge the reader 
to refer to Heavens et al. (2006) and Taylor et al. (2007) for a complete and in-depth introduction 
to the methods. In Section \ref{Methodology} we also discuss dark energy parameter prediction. 
In Section \ref{Primary Systematic Effects} we introduce the primary systematic effects considered, 
and the parameterisations used. The 
potential bias in dark energy parameters due to each systematic effect is presented in Section 
\ref{Bias in Dark Energy Parameters}. A marginalisation over systematic parameters in presented in 
Section \ref{Marginalising over systematic parameters}. We conclude and recap with a discussion 
in Section \ref{Discussion}. For any technical details concerning the 3D weak lensing methods and 
systematic parameters see Appendix A and B. 

\section{Methodology}
\label{Methodology}
To analytically investigate the systematic effect on dark energy parameter estimation the 
Fisher matrix methodology will be used throughout this paper. In the case of Gaussian-distributed 
data, which will
assumed throughout, the Fisher matrix is given for a set
%\footnote{Note that 
%the set can be arranged as a vector $\underline\Phi$ in which case the union of two parameter sets 
%is $\underline\Phi=(\underline\theta,\underline\psi)=\underline\theta+\underline\psi$.} 
of parameters $\Phib$ by (Tegmark, Taylor \& Heavens, 1997; Jungman et al., 1996; Fisher, 1935)
\be
\label{fishe}
F_{ij}=\frac{1}{2}{\rm Tr}[C^{-1}C,_iC^{-1}C,_j+\mu,_iC^{-1}\mu^T,_j+\mu,_jC^{-1}\mu^T,_i],
\ee
where $C=S+N$ is the covariance matrix of the method, 
a sum of signal $S$ and noise $N$ terms, and $\mu$ is the mean of the signal. Commas  
denote derivatives with respect to a parameter, $C,_i\equiv dC/d\Phi_i$.

There are a number of methods which employ shear and redshift information in order to constrain 
cosmological parameters. The most widely studied is weak lensing tomography, which is an  
extension of 2D cosmic shear to multiple redshift bins, and as such occasionally referred 
to as a 2D+1 method. There are various incarnations of the
tomographic technique. Hu (1999) and Takada \& White (2004) include all correlations between 
redshift bins. Takada \& Jain (2004) introduced bispectrum tomography. Jain \& Taylor (2003) 
introduced the concept of taking ratios of tomographic bins which was extended to 
Cross Correlation Cosmography (CCC) by Bernstein \& Jain (2004). The effect of systematic errors 
on weak lensing tomography has also been extensively studied by 
for example Ma, Hu \& Huterer (2006), Huterer et al. (2006), Ishak (2005), King \& Schneider (2002), 
Bridle \& King (2007), Amara \& Refregier (2007). However these papers generally 
only consider each systematic individually here we will combine the effect of the 
three systematics considered. Furthermore 
this paper will focus on two different 3D weak lensing methods. 

\subsection{Shear-Ratio}
\label{Shear-ratio method}
The shear-ratio method (Jain \& Taylor, 2003; Taylor et al., 2007) takes
the ratio of the average tangential shear $\gamma^t$ around galaxy clusters in an 
exhaustive set of various redshift bin pairs 
\be 
R_{ij}=\frac{\gamma^t_j}{\gamma^t_i}.
\ee
By taking this ratio any dependence on the mass, or shape, of the lensing cluster 
drops out resulting in a signal that 
only depends on the geometry of the observer-lens-source system. 
The cosmological parameters that can be constrained 
are therefore only those that affect the geometry of the Universe: $\Omega_m$, $\Omega_{de}$, 
$w_0$ and $w_a$. We parameterise 
$w(z)=w_0+w_a z/(1+z)=w_0+w_a(1-a)$ (Chevallier and Polarski, 2001) 
and we do not assume spatial flatness. 
The redshift range is maximally and exhaustively binned such that the bin width at any 
redshift is equal to the photometric redshift error at that redshift. The leakage or scatter 
of galaxies 
between bins due to the photometric redshift uncertainty is fully taken into account. The 
abundance of clusters as a function of mass and redshift is modelled using the halo model. To 
constrain cosmological parameters (Kitching et al., 2007) the ratio of tangential shears behind a
cluster is measured directly and fitted by varying the theoretical shear-ratio estimate. The 
Fisher matrix is therefore calculated by varying the mean in equation (\ref{fishe}).
 
\subsection{3D Cosmic Shear}
\label{3D Cosmic Shear}
3D cosmic shear (Heavens, 2003; Heavens et al., 2006) requires no binning in 
redshift, describing the entire 3D shear field using a 3D spherical harmonic expansion 
for a small angle survey. 
The transform coefficients for a given set of azimuthal $\ell$, and radial $k$ 
[$h$Mpc$^{-1}$], wave numbers are given by summing over all galaxies $g$; 
\be
\hat\gamma(k,\ell)=
\sum_g \gamma^g kj_{\ell}(k r^g){\rm e}^{-i\ell.\theta^g},
\ee
following the conventions of Castro et al. (2005), and we assume a flat sky approximation. 
Since the mean signal of the 
coefficients is zero the covariance is varied until it matches that of the data 
(Kitching et al., 2007). The 3D cosmic shear covariance 
depends on the matter power spectrum as well as the lensing geometry, so the total parameter set
that can be constrained is:
$\Omega_m$, $\Omega_{de}$, $\Omega_b$, $h$, $\sigma_8$, $w_0$, $w_a$, $n_s$ and 
the running of the spectral index $\alpha_n$. Again we do not assume spatial flatness, 
and we have neglected the effect of massive neutrinos (Kitching et al, 2008b address the 
potential of 3D cosmic shear in constraining massive neutrinos). 
We use an $\ell_{\rm max}=5000$ and a $k_{\rm max}=1.5$ Mpc$^{-1}$ 
and use the same assumptions presented in Heavens et al. (2006). 

\subsection{Dark Energy Predictions}
\label{Dark Energy Predictions}
If one has a set of cosmological parameters $\btheta$ and a set of parameters describing 
a systematic effect $\bpsi$ then the total parameter set is given by 
$\Phib=(\btheta,\bpsi)$.
If extra parameters are added to the signal part of a method (either the mean or the covariance) 
the new Fisher matrix becomes a 
combination of the cosmological Fisher matrix $F^{\theta\theta}$, the derivatives of the likelihood 
with respect to the cosmological parameters and the systematic parameter $F^{\theta\psi}$ and the 
systematic parameters with themselves $F^{\psi\psi}$. So that for the total parameter set 
$\Phib$ the Fisher matrix is defined as
\be
\label{fishf}
F^{\Phi\Phi}=
\left( \begin{array}{cc}
 F^{\theta\theta} & F^{\theta\psi}  \\
F^{\psi\theta}   & F^{\psi\psi} \\
  \end{array}\right).
\ee

For all results shown we include a predicted 14-month Planck prior (described in Taylor et al., 
2007) for which we use the parameter set $\Omega_m$, $\Omega_{de}$, $h$, $\sigma_8$, 
$\Omega_b$, $w_0$, $w_a$, $n_s$, $\alpha_n$, optical depth $\tau$ and the 
tensor to scalar ratio $r$. All dark energy constraints quoted are fully marginalised over all
other parameters. We assume a $\Lambda$CDM (best fit WMAP3; Spergel et al., 2007) 
fiducial cosmology throughout with 
$\Omega_m=0.27$, $\Omega_{de}=0.73$, $h=0.71$, $\sigma_8=0.80$, $\Omega_b=0.04$, $w_0=-1.0$, 
$w_a=0.0$, $n_s=1.0$, $\alpha_n=0.0$, $\tau=0.09$ and $r=0.01$.

The performance of a particular survey in terms of its ability to measure the dark energy 
equation of state is commonly quantified in terms of a dark energy `Figure of Merit' (FoM) 
(Dark Energy Task Force Report; Albrecht et al., 2006). 
Parameterising the dark energy equation of state's 
redshift behaviour using $w(z)=w_0+w_a(1-a)$ there exists a pivot redshift $z_p$, at which 
the constraints using this parameterisation minimise, and a corresponding error on $w(z)$ at 
that redshift $\Delta w(z_p)$, this corresponds to rewriting the equation of state as 
$w(z)=w_p+w_a(a_p-a)$. The FoM is given by the reciprocal of the 
area of the $1$-$\sigma$ (two parameter) ellipse at the pivot redshift 
\be 
\label{FoM}
{\rm FoM}=\frac{1}{\Delta w_a \Delta w(z_p)}.
\ee
Note that the Dark Energy Task Force uses the inverse area of the 2-$\sigma$ ellipse. 
The results here will show the effect on the FoM for a fiducial survey design, the 
parameters of this survey are shown in Table \ref{surveys}. We will 
take this to be a $20,000$ square degree survey with a median redshift of $z_m=0.9$ and a surface 
number density of $n_0=35$ galaxies per square arcminute. We assume a fiducial 
redshift error of $\sigma_z(z)=0.025(1+z)$ and an intrinsic ellipticity dispersion of 
$\sigma_{\epsilon}=0.247$ per shear component (note this differs by a factor of $\sqrt{2}$ from the dispersion on the total shear), this is similar to a 
next generation DUNE-type experiment (the DUNE surface number density though is slightly higher). We also consider a Pan-STARRS-1 (PS1) type experiment, which 
has substantially different survey parameters that any results presented can be tested 
for consistency over survey design, this will be done in Section \ref{Combined constraints}.

The 
predicted baseline constraints for the two surveys, for the two 3D weak lensing methods 
and the combined constraints (see Section \ref{Combined constraints}) are shown in
Table \ref{baseline}. These constraints are in agreement with other predictions for DUNE 
(for example Amara \& Refregier, 2007) and with the Dark Energy Task Foce report
(Albrecht et al., 2006) for a Stage III/IV weak lensing space-based mission for DUNE and a 
Stage II/III ground-based weak lensing survey for Pan-STARRS.

\begin{table}
\begin{center}
\begin{tabular}{|l|c|c|}
\hline
Survey&DUNE (fiducial)&PS1\\
\hline
Area/sqdeg&$20,000$&$20,000$\\
$z_{\rm median}$&$0.90$&$0.75$\\
$n_0$/sqarcmin&$35$&$5$\\
$\sigma_z(z)/(1+z)$&$0.025$&$0.05$\\
$\sigma_{\epsilon}$&$0.25$&$0.30$\\ 
\hline
\end{tabular}
\caption{The parameters describing the surveys investigated. Throughout the fiducial survey 
  will use the DUNE survey parameters.}
\label{surveys}
\end{center}
\end{table}

\begin{table}
\begin{center}
\begin{tabular}{|l|l|l|l|}
\hline
DUNE(fiducial)&Shear-Ratio&3D Cosmic Shear&Combined\\
\hline
$\Delta w_0$&$0.032$&$0.031$&$0.021$\\
$\Delta w_a$&$0.177$&$0.133$&$0.108$\\
$z_p$&$0.248$ &$0.295$&$0.245$\\
$\Delta w(z_p)$&$0.015$&$0.016$&$0.015$ \\
FoM&$365$&$475$&$601$\\
\hline
\\
\hline
PS1&Shear-Ratio&3D Cosmic Shear&Combined\\
\hline
$\Delta w_0$&$0.073$&$0.074$&$0.029$\\
$\Delta w_a$&$0.322$&$0.288$&$0.138$\\
$z_p$&$0.310$&$0.169$&$0.178$\\
$\Delta w(z_p)$&$0.028$&$0.025$&$0.020$ \\
FoM&$110$&$140$&$363$\\
\hline
\end{tabular}
\caption{The baseline constraints for the two 3D weak lensing 
methods for the survey designs considered (described in 
Section \ref{Dark Energy Predictions}). 
The results shown all include a Planck prior. The FoM is given by equation (\ref{FoM}). The 
combined constraints are the predicted constraints from combining the two methods, discussed 
in Section \ref{Combined constraints}.}
\label{baseline}
\end{center}
\end{table}

\section{Primary Systematic Effects}
\label{Primary Systematic Effects}
In this Section we will introduce simple parameterisations that will be used to investigate three
systematic effects that may have the largest effect on dark energy parameter estimation; 
photometric redshift estimates, shear distortions and intrinsic alignments. 

\subsection{Photometric redshift uncertainities}
\label{Photometric redshift uncertainities}
Methods which can constrain the dark energy equation of state necessarily need to include some 
redshift information, since dark energy is an accelerating effect changing the expansion history
and growth of structure over time (redshift). Note that we will be concerned soley with 
systematic effects inherent in multi-band photometric surveys. Experiments such as the 
Square Kilometer Array (SKA; Blake et al., 2004) 
will not have photometric redshift uncertainties, so that any systematic effect related to
photometric redshifts could be ignored, however this may be at the expense of other 
potential systematics specific to measuring shear using radio data (Chang, Refregier \& 
Helfand, 2004). 
 
For wide field and relatively deep surveys, consisting of $10^7$ to $10^9$ galaxies,
which can be used to constrain cosmological 
parameters to the percent level, spectroscopic redshifts are currently unfeasible and 
so photometric redshifts must be utilised. There are many techniques that can be used to gain 
a redshift estimate from photometric data for example neural-networks ({\tt ANNz}; 
Collister \& Lahav, 2004), chi-squared fitting 
({\tt Hyper-Z}; Bolzonella, Miralles \& Pell\'o, 2000) and 
Bayesian estimation (Benitez, 2000; Edmondson et al., 2006; Feldman et al., 2006). 
However, due to the inherent limitations of the photometric technique, all methods 
result in an error on a given redshift estimation and a scatter in the relation between the true 
(spectroscopic) redshift and the photometric redshift. The success of a photometric redshift 
estimation procedure is most commonly represented as a scattered distribution in the 
spectroscopic-photometric ($z_{\rm s}$, $z_{\rm ph}$) plane. 

In order to retain an independent prediction for the 
systematic effects resulting from photometric redshift estimation we will present a generic 
description of the ($z_{\rm s}, z_{\rm ph}$) plane and marginalise over
all parameters that are used in this description for each 3D weak lensing method. 
We assume a Gaussian probability distribution in redshift with a fraction of outliers, also with 
a Gaussian distribution, so that the resulting distribution is a sum of two Gaussians
\ba
p(z|z_p)&=&\frac{1-f_{\rm out}}{\sqrt{2\pi}\sigma_z(z_p)}
{\rm e}^{-(z-c_{\rm cal}z_{\rm ph}+z_{\rm bias})^2/2\sigma^2_z(z_p)}\nn
&+&\frac{f_{\rm out}}{\sqrt{2\pi}\sigma^{\rm o}_z(z_p)}
{\rm e}^{-(z-c^{\rm o}_{\rm cal}z_{\rm ph}+z^{\rm o}_{\rm bias})^2/2[\sigma^{\rm o}_z(z_p)]^2}.
\ea
Here we have assumed that the photo-z distribution is calibrated, but imperfectly, so that the median
spectroscopic redshift distribution is biased and inclined so that it lies along a line
\be 
\label{e1}
z_{\rm s}=c_{\rm cal}z_{\rm ph}+z_{\rm bias},
\ee
where $z_{\rm bias}$ is some bias and $c_{\rm cal}$ is a calibration. A value 
$c_{\rm cal}=1$ would mean that
the photometric redshift estimation is perfectly calibrated to a spectroscopic sample. 
The redshift error $\sigma_z(z)$ is assumed to be unknown and the 
distribution is also assumed to lie between some photometric redshift range $z_{\rm range}$.  
\begin{figure}
 \psfig{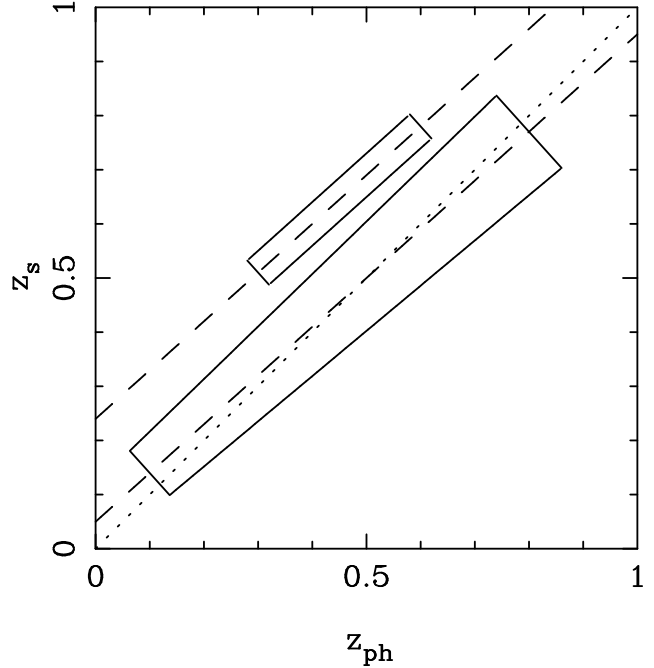}
 \caption{A simple description of the ($z_{\rm s},z_{\rm ph}$) plane. The dotted line is the 
   $z_{\rm s}=z_{\rm ph}$ around which unbiased, and perfectly calibrated, photometric redshifts 
   would lie. The dashed lines are the $z_{\rm s}=c_{\rm cal}z_{\rm ph}+z_{\rm bias}$ lines
   for the main sample (larger block) and an outlying sample (smaller block). The fiducial 
   redshift error behaviour is $\sigma_z(z)=0.025(1+z)$.}
 \label{zszp}
\end{figure}
We also assume a fraction $f_{\rm out}$ of outlying galaxies in the sample centered at a redshift 
$z^{\rm o}_{\rm mean}$, inclined on a slope described by $z^{\rm o}_{\rm bias}$ and 
$c^{\rm o}_{\rm cal}$, covering a redshift range 
$z^{\rm o}_{\rm range}$. The outlying sample's redshift error is also unknown $\sigma^{\rm o}_z(z)$. 
Note that we do not include outliers that have low spectroscopic redshifts but a broad range in 
estimated photometric redshift. Our analysis may be pessimistic in this case since by including 
such a sample some redshift biasing effects may cancel-out in this analytic approximation.  

The total set of free parameters associated with this simple 
photometric redshift parameterisation is (with fiducial values): 
$z_{\rm bias}=0.0$, $c_{\rm cal}=1.0$, $\sigma_z(z)=0.025(1+z)$, $z_{\rm range}=1.50$, 
$f_{\rm out}=0.05$, $z^{\rm o}_{\rm mean}=0.5$, $z^{\rm o}_{\rm bias}=0.1$, 
$c^{\rm o}_{\rm cal}=1.0$, $z^{\rm o}_{\rm range}=0.5$, $\sigma^{\rm o}_z(z)=0.05$. 
The fiducial values are taken to be representative of 
the photometric redshift techniques available. The results presented have been tested
against the fiducial values and there is less than a $2\%$ change in the FoM results 
for a $10\%$ change in the fiducial values. 

\subsection{Distortion of the shear}
\label{Distortion of the shear}
We analytically investigate the potential systematic effects of shear  
distortions by introducing the following parameterisation of the shear
\be
\label{sdist}
\gamma\rightarrow A_{\gamma}{\rm e}^{-i2\phi}\gamma+\gamma_{\rm bias}
\ee
where we have included an unknown rotation of the shear field 
${\phi}$, an uncertainty in the shear measurement $A_{\gamma}$ and a possible bias in the shear 
$\gamma_{\rm bias}$. This parameterisation, 
albeit simplified, should give a good idea of the effects to be expected. 
This is similar to the Shear TEsting Programme (STEP; Heymans et al., 2006a) parameterisation
where we have introduced an extra term due to rotation. The STEP parameters are 
$m= 1 - A_{\gamma}$ and $c=\gamma_{\rm bias}$ where $\phi = 0$ for all values of $m$ and $c$ . 
We take 
fiducial values of $A_{\gamma}=1.001$ and $\gamma_{\rm bias}=0.001$, 
motivated by the 
STEP results and we use a fiducial value $\phi=0.001$. 
This parameterisation represents, in an analytical and generic way, 
the combined effects of CCD distortions, 
instrument effects and inaccuracies in the shear measurement 
procedure due to the shear measurement method or image pipeline. 
We acknowledge that this is a simplistic way to include such a variety of complex
effects, but in the analytic spirit of this paper this simple model should yield a benchmark
upon which to gauge how much cosmological parameter constraints can be degraded by such effects. 

Note that we do not name this `image distortion' since the effects that we aim to 
parameterise may come from inaccuracies in the shear measurement method or instrument effects
as well as distortions and glitches in the images themselves. 

\subsection{Intrinsic alignment effects}
\label{Intrinsic alignment effects}
Intrinsic alignment effects include any non-cosmological, contaminating, source of shear. The 
spurious lensing signal from the tidal alignment of close pairs of galaxies (Heavens, 
Refregier \& Heymans, 2000; Crittenden et al., 2000; 
Brown et al., 2002; 
Catelan, et al., 2001; Heymans \& Heavens, 2003; King \& Schneider, 2003), called the 
intrinsic-intrinsic (II) term,
can potentially be removed by
ignoring the contribution to the shear covariance from close pairs of galaxies 
to any weak lensing statistic. 
Hirata \& Seljak (2004) identified a more subtle source of intrinsic shear due to the alignment 
of a background galaxies shear with the tidal field of foreground galaxies, called the shear-intrinsic 
(GI) term. Note that we will refer to GI and II, the combination of both intrinsic alignment effects 
will be refered to as IA (Intrinsic Alignments).

We use the fitting formulae given by the numerical simulations of 
Heymans et al. (2006) to the II and GI 
term. The GI fitting formula parameterises the effect using an 
amplitude $A_{\rm GI}$ and a scale dependence $\theta_0$ (Heymans et al., 2006; equation 12)
\be
\label{e2}
\langle \gamma(z_s)e^*(z_l)\rangle_{\theta}=E(z_s,z_l)\frac{A_{\rm GI}}{\theta+\theta_o}.
\ee
The free parameters were fitted using n-body simulations. $E(z_s,z_l)=D(z_l) D(z_s-z_l)/D(z_s)$ 
is the lensing efficiency of the lens-source pair, where $D$ are angular diameter distances. Note that
$A_{\rm GI}$ has units of $h$Mpc$^{-1}$ arcmins (see Table \ref{fidvals}).

Similarly we use the fitting formula given by the numerical 
simulations of Heymans et al. (2006) to the II term. 
This fitting formula parameterises the effect using an amplitude $A_{\rm II}$ only 
(Heymans et al., 2006; equation 6)
\be
\label{e2a}
\langle e(z_a)e^*(z_b)\rangle_{\theta}=\frac{A_{\rm II}}{1+(r_{ab}/B_{\rm II})^2}.
\ee
$B_{\rm II}=1 h^{-1}$Mpc is assumed to be a fixed parameter, 
where $r_{ab}$ is the comoving distance between two galaxies.

\subsection{Overview of the effects of the systematic parameters}
\label{Overview of the systematic parameters}
For a full description of how the paramerisation of the systematic 
effects are included in each of the 3D weak lensing methods see Appendix A and B. 

To summarise, the possible effect that a systematic can 
have on a method is that it can either introduce extra parameters, or add
an extra covariance. If the method involved is one in which the covariance is varied to match the 
data (as in the 3D cosmic shear case) then an extra covariance can also lead to extra parameters. 
If the method varies the mean signal (as in the shear-ratio method) then any
extra covariance adds noise and no extra parameters. Whether extra parameters are added as a result of 
the covariance is a property of the method not the systematic effect. Table \ref{tickcross} summarises
the effect on each 3D weak lensing method of including the three primary systematics. It can be 
seen that 3D cosmic shear has many more extra parameters, and that the shear-ratio method, 
whilst having fewer extra systematic parameters, suffers from more extra covariance terms. 

We note in passing that there exist three types of systematic effect that we define below. 
We assume covariance matrix, $C=S+N$, is a sum of signal $S$ and noise $N$. 
\begin{itemize}
\item
Type I: systematic $\psi$ alters signal (mean of the signal or signal covariance; depending on
the method) but not the noise $C=S(\psi)+N$.  A strong (and correct) prior on $\psi$ 
removes systematic.
\item
Type II: systematic $\psi$ adds to the covariance but not the signal $C=S+N+C^S(\psi)$. 
Strong (and correct) prior on $\psi$ does not
remove effect of systematic (errors on parameters are still increased by $\psi$
even if it is known). $\psi$ can be marginalised over if the method varies the covariance to match 
the data. 
\item 
Type III: Types I+II; which adds both extra parameters to the signal 
and adds a covariance $C=S(\psi)+N+C^S(\psi)$. 
A strong (and correct) prior on $\psi$ removes the dependence of the signal 
on the systematic parameters but does not necessarily remove the extra covariance.
\end{itemize}
If we have a Type I systematic effect, i.e. $C=S(\psi)+N$ which does not add any further 
covariance then the entire systematic effect 
can be encapsulated by marginalising over the total parameter set, which is 
equivalent to measuring (self-calibrating) the systematic parameters from the data itself. 
In this way any 
degeneracies between the systematic parameters and cosmological parameters are taken 
into account. In this case the effect of the systematics could potentially be 
alleviated by including extra information (a prior) on the systematic parameters. Note that one 
has to rely on any parameterisation used being a good description, the inherent 
danger is that the form is not accurate enough. 

A Type II systematic effect will add an extra covariance, $C=S+N+C^S(\psi)$, which 
may also contain extra 
systematic parameters, and may also be dependent on the cosmological parameter set
(or some subset). If the method used varies the covariance, as opposed to the mean signal, to 
constrain cosmological parameters then any extra covariance terms 
are marginalised over as before. However 
even if extra information is included on the systematic parameters there may  
remain an extra covariance term. For example if the covariance $C=S+N+AC^S(\psi)$ where 
$A$ is a new (nuisance) parameter the extra covariance will only be eliminated if $A=0$ not if the 
error on $A$ is zero $\Delta A=0$. If the dependence of $C^S(\psi)$ 
on the cosmological parameter set is small this will effectively 
add an extra noise term to the covariance. 

This classification is manifest in Table \ref{tickcross} where we classify each of the systematic 
effects considered in this paper for each 3D weak lensing method. Note that the 
type of systematic effect depends on the method not the systematic effect itself; the shear-ratio test
varies the mean to match to the data whereas 3D cosmic shear varies the covariance to match the data. 
Table \ref{tickcross} shows 
those systematic effects that alter the signal (Type I), those that add 
an extra covariance (Type II). This highlights the danger of assuming that 
\emph{all} systematic effects may be removed in future by marginalising over extra parameters. 

\begin{table}
\begin{center}
\begin{tabular}[c]{|l|l|l|l|}
\hline
Shear-Ratio&Signal&Cov.&Type\\
\hline
pz         &$\surd$ &$\times$&I\\        
$\gamma_S$ &$\times$&$\surd$&II\\       
IA         &$\times$&$\surd$&II\\        
\hline
\hline
3D Cosmic Shear&Signal&Cov.&Type\\
\hline
pz&$\surd$ &$\times$&I\\
$\gamma_S$&$\surd$ &$\times$&I\\
IA&$\times$&$\surd$&II\\
\hline
\end{tabular}
\caption{A summary of the effect of each systematic on the two 3D weak lensing methods. A $\surd$ 
or $\times$ indicates that either a systematic effects the signal or  
adds an extra covariance term(s). 
`pz' refers to photometric redshift systematics, `$\gamma_S$' to shear distortion and 
`IA' to intrinsic alignments. The systematics are categorised into Types in Section 
\ref{Overview of the systematic parameters}}
\label{tickcross}
\end{center}
\end{table}

For a list of all the extra systematic parameters used,
and their fiducial values see Table \ref{fidvals}. 
The photometric redshift values are interpolated from the literature, 
for example Abdalla et al. (2007), 
though no specific reference gives definitive values. The shear distortion values are taken 
from the STEP papers (Heymans et al., 2006a; Massey et al., 2007). The intrinsic alignment values 
are taken from an n-body simulation of the intrinsic alignment effects, Heymans et al. (2006). 
The GI terms
have different values for the two methods since the shear-ratio method uses tangential shear
whereas the 3D cosmic shear method uses the $\gamma_1$ and $\gamma_2$ components of shear directly. 
Heymans et al. (2006) give different intrinsic alignment systematic parameter values 
for $\gamma_t$, $\gamma_{\times}$ and $\gamma$, 
we take the most likely values of the intrinsic alignment parameters.  
\begin{table}
\begin{center}
\begin{tabular}[c]{|l|l|}
\hline
Extra Parameter&Fiducial Value\\
\hline
\hline
{\bf Photo-z}\\
$z_{\rm bias}$&$0.0$\\
$c_{\rm calibration}$&$1.0$\\
$\sigma_{z}/(1+z)$&$0.025$\\
$z_{\rm range}$&$1.5$\\
\hline 
{\bf Outliers}\\
$f_{\rm out}$&$0.05$\\
$z^0_{\rm bias}$&$0.1$\\
$c^0_{\rm calibration}$&$1.0$\\
$\sigma^0_{z}$&$0.05$\\
$z^0_{\rm mean}$&$0.5$\\
$z^0_{\rm range}$&$0.5$\\
\hline
\hline
{\bf Shear Distortion}\\
$A_{\gamma}$&$1.001$\\
$\phi$&$0.001$\\
$\gamma_{\rm bias}$&$0.001$\\
\hline
\hline
{\bf Shear-Ratio IA ($\gamma_t$)}\\
$A_{\rm GI}/(h$Mpc$^{-1}$ arcmin)&$-0.92\times 10^{-8}$\\
$\theta_{0}$/arcmin&$1.32$\\
$A_{\rm II}$&$0.45\times 10^{-3}$\\
\hline
\hline
{\bf 3D Cosmic Shear IA ($\gamma$)}\\
$A_{\rm GI}/(h$Mpc$^{-1}$ arcmin)&$-1.26\times 10^{-7}$\\
$\theta_{0}$/arcmin&$0.90$\\
$A_{\rm II}$&$0.47\times 10^{-3}$\\
\hline
\end{tabular}
\caption{A list of the extra systematic parameters used and the Fiducial values adopted.}
\label{fidvals}
\end{center}
\end{table}

\section{Bias in Dark Energy Parameters}
\label{Bias in Dark Energy Parameters}
Instead of assuming an extra systematic parameter is measured from the data 
(and marginalising over its value) an approach can be adopted 
in which the extra parameter's effect, which will be a bias, 
on the cosmological parameters is estimated whilst the extra parameter's value 
is assumed to be fixed. This bias occurs due to assuming the parameter to be fixed, 
but possibly biased by an unknown amount. The cosmological parameter constraints 
will be smaller than if the extra parameter is marginalised over 
at the expense of this bias. 

In Taylor et al. (2007) it was shown how 
to calculate such biases using a Fisher matrix approach. The linear bias
in a measured parameter $\delta\theta_i$ 
due to a bias in a fixed model (systematic) parameter $\delta\psi_j$ is given by 
\be
\label{biaseq}
\delta\theta_i=-(F^{\theta\theta})^{-1}_{ik}F^{\theta\psi}_{kj}\delta\psi_j.
\ee
$F^{\theta\theta}$ is the Fisher matrix of measured (cosmological) parameters and 
$F^{\theta\psi}$ is a matrix of derivatives with respect to parameters assumed to be fixed and 
those assumed be be measured (see equation \ref{fishf}). We also define $C_{\psi}$ where 
\be
\label{biaseq2}
\delta\theta=-C_{\psi}\delta\psi
\ee 
which characterises a systematic parameters biasing effect on a cosmological parameter. 

Note that throughout we use $\Delta\theta$ to represent the marginal error on a parameter and 
$\delta\theta$ to represent the offset in the maximum likelihood value of a parameter. 

To encapsulate the biasing effect in the dark energy ($w[z_p]$, $w_a$) parameter space we introduce 
a Bias Figure of Merit (BFoM) which is defined as 
\be
\label{eBFoM}
{\rm BFoM}=\frac{1}{|\delta w(z_p)\delta w_a|}
\ee
where $\delta w(z_p)$ and $\delta w_a$ are the biases in the pivot redshift value and the value of 
$w_a$ due to assuming a systematic parameter to be fixed. One would wish to maximise the BFoM, its 
value tending to infinity for zero bias. A desirable bias of less than $\sim 0.01$ in both 
$w(z_p)$ and $w_a$ results in a BFoM$=10,000$ whereas a poor bias would be of order BFoM$\ls 100$. 
We stress that this is, in analogy with the FoM, a diagnostic tool only so that a 
conceptual understanding of the relative effect of the systematic parameters can be gained. It does not
represent any fundamental aspect of the likelihood surface and is of course contingent on both 
the parameterisation of $w(z)$ used and in this case the assumption of Gaussianity implicit in the 
Fisher matrix formalism. Furthermore it has 
the potential, as does the FoM to become artificially dominated by a good result on one parameter 
masking the poor result of the other. 
 
\begin{table}
\begin{center}
\begin{tabular}{|l|l|l|l|}
\hline
Extra Parameter&Shear-Ratio\\
&$\delta w(z_p)$&$\delta w_a$&BFoM\\
\hline
$z_{\rm bias}$                 &$  +0.5577 $&$       -3.0418   $&$    0.5899$\\
$c_{\rm calibration}$          &$  +0.4026 $&$      -0.0275 $&$   90.055$\\
$\sigma_{z}$ &$  +0.0530 $&$   -2.4449  $&$     7.7140$\\
$z_{\rm range}              $&$ -0.0184$&$ -0.1520  $&$       364.64$\\
$f_{\rm out}           $&$  +0.0047$&$   -0.3236 $&$       659.12$\\
$z^0_{\rm bias}              $&$  +0.0037$&$   -0.1573  $&$      1740.1$\\
$c^0_{\rm calibration}           $&$  +0.0018 $&$  -0.0560 $&$   9719.8$\\
 $\sigma^0_{z}     $&$ -0.0044$&$   +0.0539 $&$   4249.7$\\
$z^0_{\rm mean}         $&$  +0.0019$&$  -0.0526$&$   9997.2$\\
$z^0_{\rm range}       $&$  +0.0018$&$   -0.0752$&$   7259.2$\\
$\phi$&$-$&$-$&$-$\\
$A_{\gamma}$&$-$&$-$&$-$\\
$\gamma_{\rm bias}$&$-$&$-$&$-$\\
$A_{\rm GI}$&$-$&$-$&$-$\\
$\theta_{0}$&$-$&$-$&$-$\\
$A_{\rm II}$&$-$&$-$&$-$\\
\hline
\hline
Extra Parameter&3D Cosmic Shear\\
&$\delta w(z_p)$&$\delta w_a$&BFoM\\
\hline
$z_{\rm bias} $          &$ -0.0548$ &$-0.6536       $ &$27.919$\\
$ c_{\rm calibration}$    &$  +0.0133$ &$ -0.1410      $ &$529.82$\\
$ \sigma_z         $   &$  +0.0015$ &$ -0.0523 $ &$12070$\\
$ z_{\rm range}        $  &$ -0.0068$ &$  +0.7471      $ &$195.05$\\
$ f_{\rm out}           $ &$ -0.0381$ &$ -0.3588      $ &$ 73.098$\\
$ z^0_{\rm bias}      $&$ -0.0029$ &$ -0.0349 $ &$ 9832.0$\\
$ c^0_{\rm calibration} $&$ -0.0015$ &$ -0.0186 $ &$ 35862$\\
$\sigma^0_z         $&$  +1.3\times 10^{-5}$ &$  +0.0004  $ &$1.7\times 10^8$\\
$ z^0_{\rm mean}      $&$ -0.0022$ &$ -0.0551 $ &$ 8278.3$\\
$ z^0_{\rm range}     $&$ -0.0007$ &$ -0.0429  $ &$ 33846$\\
$ \phi                 $&$ -6.1\times 10^{-5}$ &$ -0.0003  $ &$5.8\times 10^7$\\
$ A_{\gamma}           $&$  +0.0123$ &$  +0.0563 $ &$ 1440.3$\\
$\gamma_{\rm bias}$&$-$&$-$&$-$\\
$ A_{\rm GI}               $&$   +0.0084    $&$ +0.0492 $&$  2404.7$\\
$ \theta_0        $&$ -3.7\times 10^{-7}$&$  -2.1\times 10^{-6} $&$  1.3\times 10^{12}$\\
$ A_{\rm II}               $&$  -0.0129    $&$   -0.027    $&$   2830.5$\\
\hline
\end{tabular}
\caption{The potential bias on the most likely value of $w(z_p)$ and $w_a$ 
due to assuming that an extra 
systematic parameter is fixed, and biased by $+0.01$. The BFoM is defined in equation (\ref{eBFoM}). 
A `$-$' is shown 
where the extra parameters do not occur in the signal part of the method and as such cannot be 
marginalised over.}
\label{biastable}
\end{center}
\end{table}

%Assuming that the extra systematic parameters are fixed, but biased, 
%can lead to a potential bias
%in the most likely value of a measured cosmological parameter. 
In Table \ref{biastable} we show 
the potential bias in $w(z_p)$ and $w_a$ due to a $+ 0.01$ bias in each extra 
systematic parameter individually, for any given parameter the bias in the others 
is assumed to be zero.  
All extra covariance and noise terms due to each systematic effect are included. Note that the sign 
of the bias in $w(z_p)$ or $w_a$ is contingent on the sign of the bias in the 
systematic parameter 
considered (in this case $+0.01$), and the magnitude is dependent on the size of the bias considered. 
If the bias in the systematic parameter was larger then the bias in $w(z_p)$ or $w_a$ would be 
proportionally larger. The values given are meant to be indicative of the sensitivity of the 
3D weak lensing methods dark energy constraints to each systematic parameter.

%From equation (\ref{biaseq2}) $\delta w(z_p)=-C_{\phi}\delta\psi$ so that for a shift in a 
%systematic parameter of $+0.01$, considered here, $C_{\phi}=\delta w(z_p)/0.01$ where 
%$\delta w(z_p)$ is given in Table \ref{biastable}. The question on bias can now be inverted from 
%`what is the bias in $w(z_p)$ given a shift of $0.01$ in a systematic parameter?' to `what is the 
%required accuracy on a given parameter given that $w(z_p)$ needs to be known to $0.01$?'. 
%The answer to the second question is given by $\delta\psi=0.01/C_{\phi}=
%1\times 10^{-4}/\delta w(z_p)$.

It can be seen that a bias in the photometric redshifts parameter 
$z_{\rm bias}$ has the largest effect for both methods. 
This indicates that $z_{\rm bias}$ would need to be accurate to $1$ part in $10^{4}$ 
for the shear ratio method and $1$ part in $10^{3}$ for the 3D cosmic shear method
for the most likely value of $w(z_p)$ to be accurate to $\pm 0.01$. 

The shear-ratio method is sensitive to all the photometric redshift parameters particularly the bias
and calibration since these affect the scatter/leakage of galaxies between bins at all 
redshifts (see Appendix A). 
The method is less sensitive to the outlying fraction of galaxies since 
these have a smaller effect on the redshift distribution.

The 3D cosmic shear method is sensitive to all of the photometric redshift parameters, 
particularly the bias, calibration and redshift range.  
The sensitivity to parameters such as the bias and calibration is a result of the parameters 
affecting the estimated redshift directly and as such the weighting of the shear estimators.
The photometric redshift extra parameters add uncertainty to the photometric 
redshift probability distribution. The exact 
form of the probability distribution has an effect on dark energy parameter estimation since, 
as shown in Castro et al. (2005) and Heavens et al. (2006), the majority of the dark energy 
signal comes from $\ell\approx 1000$. Through the Bessel function 
maximum at $\ell \approx kr_{\rm max}$ this corresponds to radial modes of $k\approx 0.35$ Mpc$^{-1}$. 
Photometric redshifts damp the radial modes at intermediate and high $k$ values, at scales of 
$k\geq 2\pi h/(3000\sigma_z)$ for $\sigma_z\approx 0.05$ this corresponds to $k\geq 0.03$. 

The bias in the shear distortion parameter (approximately the STEP parameter $m$) 
has to be $\delta A_{\gamma}\leq (0.01^2/0.0123)=0.008$ 
for the bias in $\delta w_p\leq 0.01$. This is in 
approximate agreement with Amara \& Refregier (2007), 
who find the bias in their $m_0$ parameter needs to be of order $0.005$, 
though they focus on investigating the bias in the variance $\sigma_{\gamma}$.

The cosmological dependence of GI and II terms is small (the $F^{\theta\psi}_{kj}$ term 
in equation, \ref{biaseq}).
This low bias for the GI and II terms extra parameters suggests that  
the intrinsic alignment terms effectively add extra noise to the 3D cosmic shear covariance. 
In particular the scale dependence of the GI term ($\theta_0$)
has a very small potential bias since this only changes the overall normalisation of the GI term in 
a small non-linear way (see Appendix B). 

%Equation (\ref{biaseq}) shows that 
%if either the errors on the parameters $\theta_i$ are poor and/or the sensitivity to the 
%extra parameters $\psi_j$ is small then the bias will be large. For the 3D cosmic shear the 
%sensitivity ($F^{\theta\psi}_{kj}$) is smaller in comparison to the shear-ratio method. 
%The parameters which have the largest biasing effect, 
%due to a larger sensitivity, are also those parameters which should be constrained well by the data 
%and so have a small effect if they were marginalised over. 

The parameters which have a large bias are also the values which should 
be well measured by the data itself: the method is very sensitive to these parameters. 
In addition the parameters which have a very small bias should have a small 
effect on the FoM: the method is not sensitive to the parameter and 
the degeneracy between the extra parameter and the dark energy parameters is small. 
Therefore it is the parameters with intermediate values of BFoM that 
should have the largest effect when marginalising over them: the method is somewhat 
sensitive to the parameter 
of interest and there is a large degeneracy between the extra parameter and the 
dark energy parameters.

We have shown that assuming that certain systematic parameters are fixed, but biased, 
can have a large effect on dark energy parameter estimation by biasing the most likely values of 
$w(z_p)$ or $w_a$ by a large amount. This suggests that marginalising over such parameters must be 
a more reliable way of dealing with such effects. When marginalising over systematic parameters 
error bars on cosmological parameters of interest will be larger but 
the most likely value of the cosmological parameter of interest will remain unbiased. 

In practice, in order to save computational time, 
one may wish to identify which parameters could cause a bias and then only marginalise over 
those which appear to be troublesome. 
However in this paper we will continue to marginalise over all available parameters 
so that their potential effect can be monitored. 

Recently Amara \& Refregier (2007) have used the bias in cosmological parameters to explore 
how a survey could be designed by fixing the parameter accuracy needed and asking what 
bias could be tolerated that would yield such an accuracy. 
In the case of designing a survey (or photo-z method for example) investigating the
maximum potential bias that can be tolerated [for a given FoM] and then
using this information as a benchmark upon which 
to minimise the bias in the survey design (or method)
is the correct procedure. The worst case is that systematic parameters will
be biased by a large and unknown amount, and one must assume this worst case 
in order to place the most stringent constraints on design.

However in the case of assessing the potential impact of systematic effects,
on the use of 3D lensing given a survey for example, marginalising over parameters 
which have a large biases is the more prudent approach. Instead of assuming that parameters
are biased by a large unknown amounts, the parameter can be marginalised over which
takes into account the full range of potential values of a parameter, not just the
largest possible deviation from its fiducial/expected value. In this case marginalising over these
parameters is the best approach; at the expense of a larger error on the
cosmological parameters the large bias is negated and the most likely value of the cosmological 
parameter remains intact. 

\section{Marginalising over systematic parameters}
\label{Marginalising over systematic parameters}
Here we investigate the effect of marginalisation over systematic nuisance parameters. 
The systematic parameters are assumed to be extra parameters that are 
measured/calibrated directly from the data. This will not lead to a bias in any cosmological 
parameters but will increase the error through degeneracies with the extra parameters. 

We again use equation (\ref{fishf}) to construct a Fisher matrix 
containing both the cosmological and extra (nuisance) parameters. 
The resultant predicted marginalised error on a given cosmological parameter $\theta$, 
is given by 
$\Delta\theta=\sqrt{[F^{\Phi\Phi}]^{-1}_{\theta\theta}}$. 
In this way the new predicted cosmological parameter error is marginalised over the predicted 
systematic parameter constraints. This can be compared with the cosmological parameter constraint 
which does not take into account the extra marginalisation over the new (nuisance) parameters,  
$\Delta\theta=\sqrt{[F^{\theta\theta}]^{-1}_{\theta\theta}}$.

In this Section we will progressively add the primary systematic effects to the 3D weak lensing 
methods in turn and asses the impact of the systematic effects on the dark energy FoM attainable using 
each method. We will present the effect of each individual systematic parameter, so that the source 
of any reduction in the FoM can be identified, aswell as the overall reduction in the FoM due to the 
systematic effects. The results are summarised in Table \ref{newline}. 

It is important to stress that in this Section we do not assume any prior information 
on the systematic parameters, and as such the 
results presented are a \emph{worst-case} scenario. In reality each systematic effect parameter 
may have prior information 
which can only improve upon the results presented in this Section. We investigate the effect 
of prior information on the systematic parameters in Section \ref{Discussion}. 

\subsection{Shear-Ratio Method}
As shown in Appendix A and summarised in Table \ref{tickcross} 
only the photometric redshift uncertainties add extra systematic 
parameters to the shear-ratio method, the shear distortion and intrinsic alignment effects add extra 
covariance terms which in this case act as extra sources of noise.  
\\
\\
\noindent {\bf a) Photometric Redshift Systematics} 

\begin{figure}
 \psfig{figure=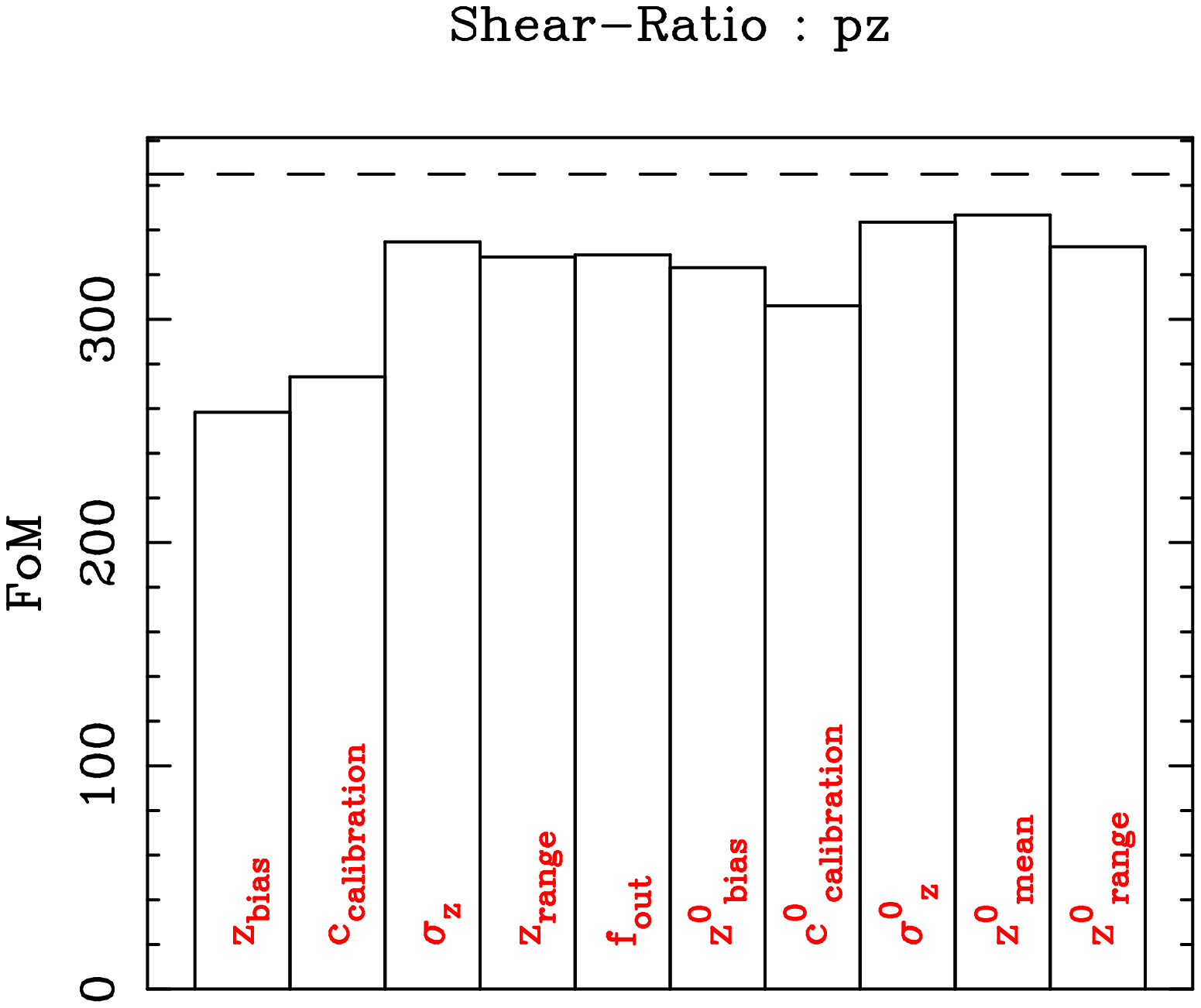,width=\columnwidth,angle=0}
 \caption{The worst-case changes in the figure of merit for the fiducial survey with each 
   of the individual photometric redshift
   systematic parameters for the shear-ratio method. 
   The dashed line shows the baseline FoM.}
 \label{geo_bar_pz}
\end{figure}

Figure \ref{geo_bar_pz} shows the reduction in the FoM for the various photometric 
redshift parameters for the shear-ratio method. 
The dark energy constraints are most sensitive to marginalising over a bias in the photometric
redshifts. This is due to the nature of the effect of $z_{\rm bias}$, in which a slight change 
affects all redshifts. The dark energy constraints are relatively insensitive to the redshift 
range over which the photometric redshifts can be used ($z_{\rm range}$) since this parameter 
does not affect all redshifts and the method is insensitive to the maximum, and minimum, 
redshifts used. The method is also relatively insensitive to the outlying population, but note 
that we use a fiducial value of $f_{\rm out}=0.05$.
The insensitivity to the majority of the photometric redshift parameters stems from the fact that 
they affect the detailed form of the photometric redshift distribution only which 
slightly changes the amount of scatter/leakage of galaxies between bins. 
The parameters which globally affect all redshifts have a larger effect on the FoM. 
When marginalising over all the photometric parameters the FoM=$163$, a factor of $2.2$ smaller 
than the baseline FoM.
\\
\\
\noindent {\bf b) Photometric Redshift Systematics \& Intrinsic Alignments}

By including the intrinsic alignment terms GI and II
as extra sources of noise the effect on the photometric redshift 
parameters is very similar to 
the case of including photometric redshift systematics alone (see Figure \ref{geo_bar_pz}). 
The maximum FoM is slightly reduced from the 
baseline FoM by the introduction of the GI and II noise terms from $364$ to $360$ 
as a result of these extra noise terms.  
%\begin{figure}
% \psfig{figure=geo_bar_pz_gi.eps,width=\columnwidth,angle=0}
% \caption{The change in the figure of merit with each of the individual photometric redshift
%   systematic parameters for the shear-ratio method with the GI and II 
%   intrinsic alignment terms included. The dashed line shows the baseline FoM.}
% \label{geo_bar_pz_gi}
%\end{figure}
The addition of the intrinsic alignment noise terms has a small effect on the dark energy FoM. 
This is due to two reasons, firstly the magnitude of the GI and II terms is at least a 
factor of $10$ times 
smaller than the large scale structure noise terms which affect this method. 
Secondly the nature of the GI and II terms 
(Appendix A, equation \ref{e4}), consisting of four positive noise contributions and four negative 
contributions each from the different bin-bin covariant combinations, 
means that some cancellation occurs reducing the effects further.
\\
\\
\noindent{\bf c) Photometric Redshift Systematics, Intrinsic Alignments \& Shear Distortion}

We now include photometric redshift, GI, II and shear distortion systematics 
in the shear-ratio method. 
The change in the FoM with the individual photometric redshift 
parameters, now including the shear distortion, GI and II noise terms, is again 
very similar to when the photometric redshift systematics alone are included 
(see Figure \ref{geo_bar_pz}). 
The maximum FoM is again reduced from the baseline FoM by the introduction of the 
GI and II noise terms 
from $364$ to $360$, the shear distortion systematic terms have a very small effect on the FoM.  
%\begin{figure}
% \psfig{figure=geo_bar_pz_phi_gi.eps,width=\columnwidth,angle=0}
% \caption{The change in the figure of merit with each of the individual photometric redshift
%   systematic parameters with the shear systematic, GI and II noise terms included. 
%   The dashed line shows the maximum FoM before marginalisation over the photometric 
%   redshift parameters. Note that this is a worst case FoM reduction for the fiducial survey.}
% \label{geo_bar_pz_phi_gi}
%\end{figure}
Marginalising over all the photometric redshift parameters and including GI, II
and shear distortion noise terms in the shear-ratio method 
the FoM$=157$ a factor $2.2$ smaller than the baseline FoM.

\subsection{3D Cosmic Shear}
All three of the primary systematic effects add extra parameters to the 3D cosmic shear method, 
and the GI and II intrinsic alignment effects add extra covariance terms. In this case, as 
we progressively add the systematic effects, the number of extra parameters will increase.
\\
\\
\noindent{\bf a) Photometric Redshift Systematics}

\begin{figure}
 \psfig{figure=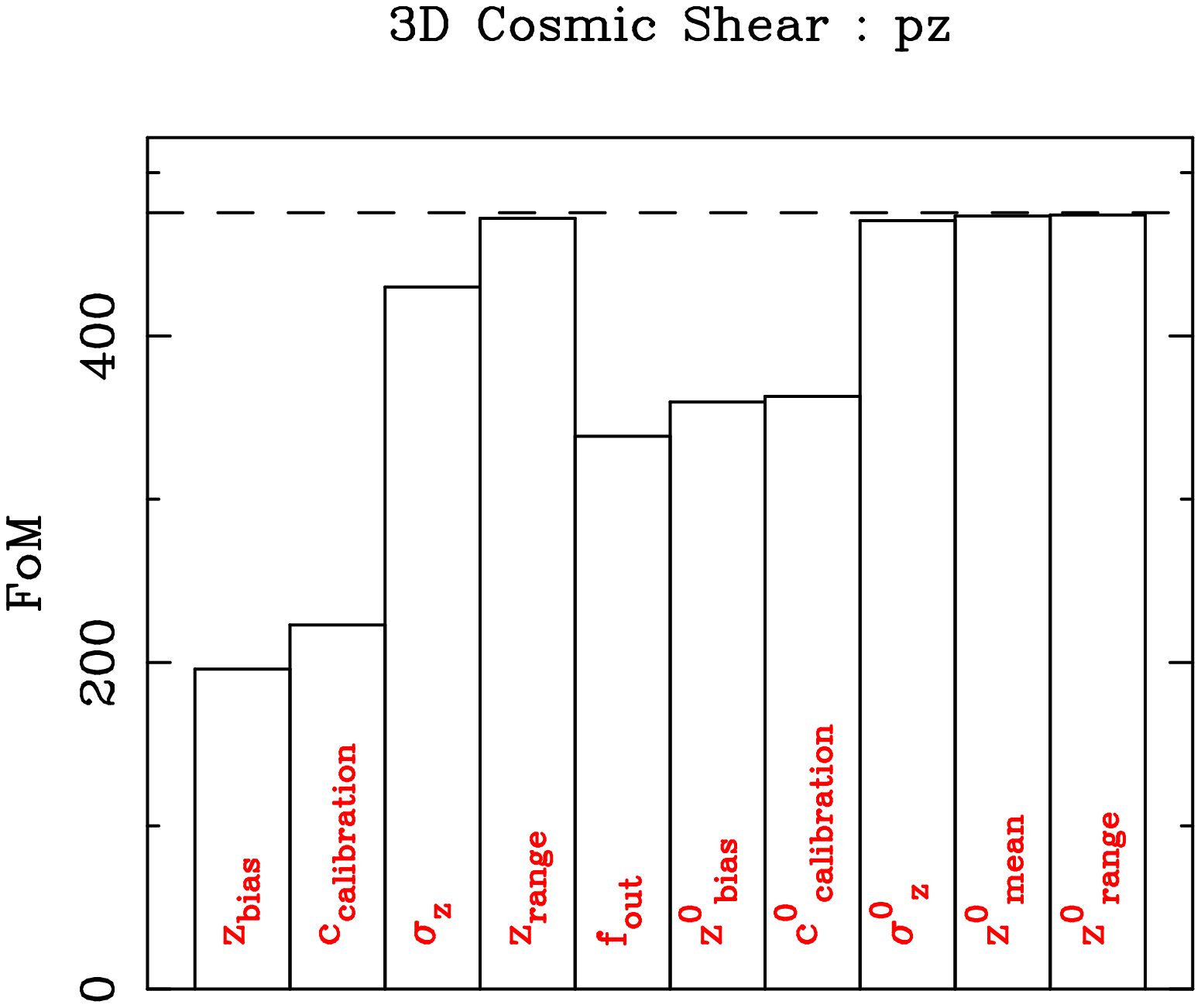,width=\columnwidth,angle=0}
 \caption{The worst-case changes in the figure of merit for the fiducial survey 
   with each of the individual photometric redshift
   systematic parameters for the 3D cosmic shear method. The dashed line shows the baseline FoM.}
 \label{spec_bar_pz}
\end{figure}

Figure \ref{spec_bar_pz} shows the change in the predicted FoM using 3D cosmic shear 
due to each of the photometric redshift systematic parameters individually. Similar to the
shear-ratio method the largest effect comes from a bias in the photometric redshifts,
this is due to the sensitivity of the method to the photometric redshift distribution, 
as discussed in Section \ref{Bias in Dark Energy Parameters}. 
When all photometric redshift parameters are marginalised over the resulting FoM is $107$, 
a factor of $4.4$ times smaller than the baseline FoM. 
\\
\\
\noindent{\bf b) Photometric Redshift Systematics \& Intrinsic Alignments}

\begin{figure}
 \psfig{figure=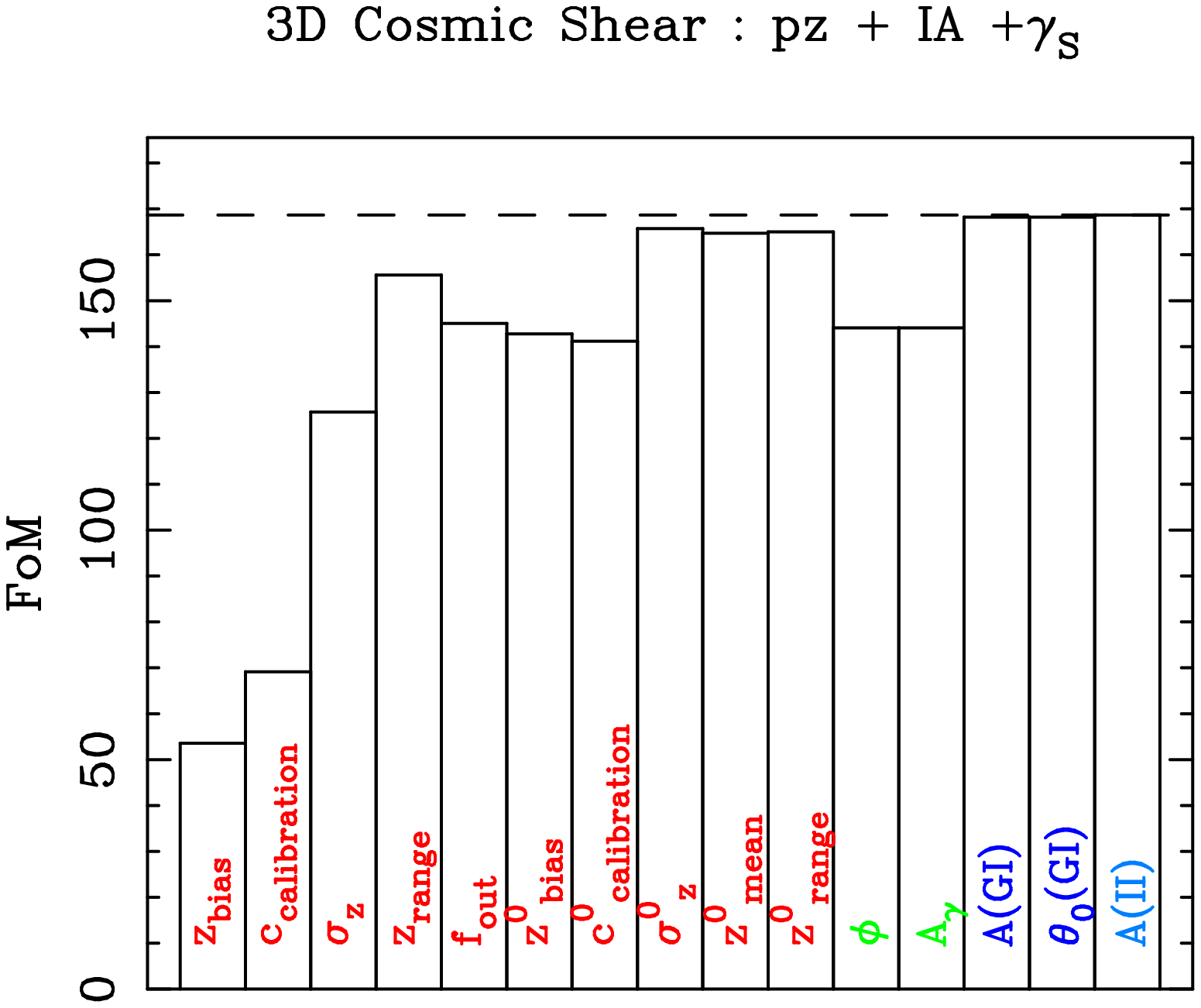,width=\columnwidth,angle=0}
 \caption{The worst-case changes in the figure of merit for the fiducial survey 
   with each of the individual 
   systematic parameters for the 3D cosmic shear method, including GI and II covariances. 
   The dashed line shows the FoM before marginalisation over extra parameters.}
 \label{spec_bar_pz_phi_gi_ii}
\end{figure}

The degradation of the FoM with each photometric redshift parameter is similar with 
the GI and II covariance terms in the 3D cosmic shear method included. 
However the maximum FoM is reduced from the 
baseline FoM by a factor of $\sim 3$ as a result of the introduction of 
extra covariance terms from $475$ to $168$ (see Figure \ref{spec_bar_pz_phi_gi_ii}).  
Marginalising over the GI and II extra parameters themselves has a small effect. This is 
due to the relatively poor cosmological dependence of the GI and II terms. 
%\begin{figure}
% \psfig{figure=spec_bar_pz_gi_ii.eps,width=\columnwidth,angle=0}
% \caption{The change in the figure of merit with each of the individual 
%   systematic parameters with GI and II covariance terms included. 
%   The dashed line shows the baseline FoM.}
% \label{spec_bar_pz_gi_ii}
%\end{figure}
Hence the GI and II terms effectively act as extra sources of noise in the covariance, the 
cosmological dependence of the covariances is small. 
%Once the GI and II terms are included there is a similar dependence on the photometric 
%redshift parameters.  
\\
\\
\noindent{\bf c)  Photometric Redshift Systematics, Intrinsic Alignments \& Shear Distortion}

As shown in Appendix B the systematic parameter $\gamma_{\rm bias}$ only has a second 
order effect 
on 3D cosmic shear and since the fiducial value of $\gamma_{\rm bias}\ll 1$ 
we only marginalise over $A_{\gamma}$ and $\phi$.
Figure \ref{spec_bar_pz_phi_gi_ii} shows the results of marginalising over each of the individual 
systematic parameters including the intrinsic alignment and shear distortion effects. It can 
be seen that the maximum FoM is again reduced due to the intrinsic alignment terms from 
$475$ to $168$. 
The individual shear distortion parameters have a small effect on the FoM as expected 
from the small bias shown in Section \ref{Bias in Dark Energy Parameters} 
implying the dark energy constraints have a small sensitivity to these parameters. 
The effects are similar for both parameters since for small $\phi$ the parameters' 
response have approximately the same functional form: 
$A_{\gamma}^2 \cos^2(2\phi)\sim (1-4\phi^2)A_{\gamma}^2$. 
Marginalising over all the systematic parameters results in a FoM$=19$ a factor of $24$ times 
smaller than the baseline FoM. 

We emphasize that the results presented in this Section are a \emph{worst-case} situation,
where the systematics are determined from the weak lensing data alone. With reasonable
priors on the systematic parameters (Section \ref{Discussion}) the situation is markedly
improved.

\section{Discussion}
\label{Discussion}
Table \ref{newline} shows the FoM, and the pivot redshift error, 
for each of the 3D weak lensing methods after each systematic 
effect is progressively added. Note that we do not display the full suite of combinations of 
systematic effects.  

The baseline constraints from the shear-ratio method for the fiducial survey design 
are shown in Table \ref{baseline}, the baseline FoM$=364$.
Since the intrinsic alignment terms and the overall shear distortion only appear in the noise 
part of the shear-ratio method, the only extra parameters to be marginalised over are
those from the photometric redshift parameterisation. 

For 3D cosmic shear the baseline FoM, using the fiducial survey design, is FoM$=475$. 
In the case of 3D 
cosmic shear the GI and II terms added extra covariances, and provided extra parameters
to marginalise over. The shear distortion systematic also provides extra parameters to be 
marginalised over. 

\begin{table}
\begin{center}
\begin{tabular}{|l|l|l|}
\hline
&Shear-Ratio&\\
&FoM&$\Delta w(z_p)$\\
\hline
baseline&$364$&$0.016$\\
pz&$163$&$0.024$\\
GI&$363$&$0.016$\\
GI+II&$363$&$0.016$\\
pz + GI&$160$&$0.024$\\
pz + GI + II&$160$&$0.024$\\
pz + $\gamma_S$&$163$&$0.024$\\
pz + GI + $\gamma_S$&$157$&$0.025$\\
pz + GI + II + $\gamma_S$ &$157$&$0.026$\\
\hline
\hline
&3D Cosmic Shear&\\
&FoM&$\Delta w(z_p)$\\
\hline
baseline&$475$&$0.015$\\
pz&$107$&$0.029$\\
GI&$300$&$0.020$\\
GI+II&$168$&$0.020$\\
pz + GI&$42$&$0.032$\\
pz + GI + II&$27$&$0.033$\\
pz + $\gamma_S$&$88$&$0.030$\\
pz + GI + $\gamma_S$&$26$&$0.033$\\
pz + GI + II + $\gamma_S$&$19$&$0.033$\\
\hline
\end{tabular}
\caption{In the worst-case with no informative priors this Table shows 
  the FoM for the two methods for the fiducial survey design as each systematic 
  effect is progressively added and all systematic parameters are marginalised over. 
  `pz' signifies that photometric redshift 
  systematics are taken into account, `GI' and `II' 
  signify that intrinsic alignment effects (GI or II) 
  are included, `$\gamma_S$' signifies that shear distortion effects are included.
  The results shown all include a Planck prior. The FoM is given by equation (\ref{FoM}).}
\label{newline}
\end{center}
\end{table}

\begin{figure}
  \centering
 \psfig{figure=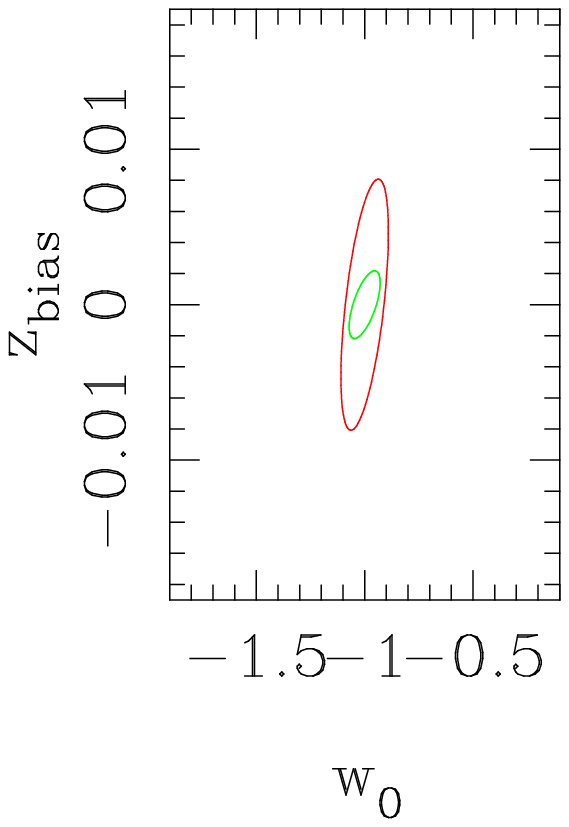,width=0.6\columnwidth,angle=0,clip=}
 \caption{The constraints from 3D cosmic shear (red/dark solid line) and the shear-ratio method 
   (green/light solid line) in the ($z_{\rm bias}$, $w_0$) 
   plane marginalising over all other photometric redshift systematic parameters, intrinsic 
   alignment and shear distortion systematics have not been included.
   A Planck prior is included.}
 \label{w0zb}
\end{figure}

The values of bias from Table \ref{biastable} and the reduction in the FoM's shown in 
Figures \ref{geo_bar_pz} to \ref{spec_bar_pz_phi_gi_ii}
highlight different aspects of the relationship between the dark energy cosmological parameters 
and the systematic parameters. 
For example it can be seen from Table \ref{biastable} that 3D cosmic shear has a smaller bias than 
the shear-ratio method for $z_{\rm bias}$. However when marginalising over 
$z_{\rm bias}$ the reduction in FoM is much larger for 3D cosmic shear than the shear ratio method. 
Figure \ref{w0zb} explains this apperent discrepancy by showing the constraints from both 3D weak 
lensing methods in the ($z_{\rm bias}$, $w_0$) plane.
The larger bias in the shear-ratio case is due to the smaller degeneracy between 
$z_{\rm bias}$ and $w_0$, a shift along the degenerate direction of the ellipse projects to a 
larger change in $w_0$ for a small change 
in $z_{\rm bias}$. However since the projection of the ellipse 
onto the $z_{\rm bias}$ axis is small the effect of marginalising over $z_{\rm bias}$  has a small 
effect on the $w_0$ constraint. In the 3D cosmic shear 
case the degeneracy is smaller, resulting in a smaller
bias, but projection onto the $z_{\rm bias}$ axis is larger resulting in 
an increase in the $w_0$ constraint when marginalising.
\\
\\
\noindent{\bf a) Photometric Redshift Systematics}

The effect of photometric redshift and shear distortion systematics is approximately 
the same for both methods. 
The similarity in the overall effects of the systematics on the methods, 
and the very different nature of the methods 
being investigated, means that some general conclusions can be made. 
For the both the shear-ratio and 3D cosmic shear methods the effect on the FoM  
from photometric redshift and shear distortion systematics results in a relative reduction 
in the FoM by a factor $\sim 2$ to $4$. 

As shown in Heavens et al. (2006) 3D cosmic shear 
is approximately $10$ times less sensitive to a photometric redshift bias
than the shear-ratio method. We find again that the bias in $w(z_p)$ in 
Table \ref{biastable} due to a bias in 
$z_{\rm bias}$ is much smaller for 3D cosmic shear than for the shear-ratio method. 
The smaller drop in the FoM when $z_{\rm bias}$ is 
marginalised over in the shear-ratio method relative to the 
3D cosmic shear method is due to this larger sensitivity as shown in Figure \ref{w0zb}. 
This larger sensitivity could be 
attributed to the binning in redshift needed for the shear-ratio method. Any quantity 
(for example a cosmological
parameter value) calculated in a given bin is calculated assuming that the galaxies are in 
that bin. If there is a bias then the derived quantity will be systematically incorrect as 
galaxies are scattered out of the bins having a large effect on the signal. 
Conversely in 3D cosmic shear a bias 
in redshift merely acts as a slightly different weighting function in redshift, a slight 
modification of the standard $j_{\ell}(r)$ weighting, so that the shear-ratio method 
is more sensitive to the redshift bias than 3D cosmic shear.
\\
\\
\noindent{\bf b) Photometric Redshift Systematics \& Intrinsic Alignments}

The effect of intrinsic alignments on the FoM 
is dependent on the 3D weak lensing method being used. 
By including photometric redshift uncertainties \emph{and} intrinsic 
alignment effects the FoM can be reduced by up to a factor of $10$, 
though we stress that this is a \emph{worst-case} scenario where no informative 
priors have been included. 

The GI and II terms have a small effect on the shear-ratio methods dark energy constraints. 
This is due the GI and II 
contributions to the covariance adding positive and negative terms which cancel to some extent
(see Appedix A). The GI term is small since we average over a small aperture 
about a cluster, using a larger 
continous area would increase this covariance. The II term is small since the intrinsic-intrinsic
correlation between any two non-nieghbouring redshift bins is small. 

The reduction in the FoM for 3D cosmic shear is 
principally due to the photometric redshift and intrinsic alignment effects. The intrinsic 
alignment effects reduce the maximum FoM by a further factor $\sim 2$, 
which is in agreement with Bridle \& King (2007). Bridle \& King (2007) 
find that for a redshift error 
of $\sigma_z(z)/(1+z)=0.05$ 
the FoM is reduced by a factor of $\sim 0.7$ by including GI alone, and by 
$\sim 0.6$ by including GI and II (from Bridle \& King, 2007; Figure 5). 
This is in comparison to what we find, shown 
in Table \ref{newline}, that the FoM is reduced by a factor of $\sim 0.6$ by including GI alone and 
by $\sim 0.3$ by including GI and II. This shows some agreement between the analyses despite the 
differences in both the 3D lensing method investigated and the intrinsic alignment parameterisations 
used. Bridle \& King (2007) showed that photometric redshifts have to be known to 
an increased accuracy when intrinsic alignment effects are included, we find a complementary result
that when marginalising over photometric redshift parameters including intrinsic alignments 
can further reduce the FoM by a factor of $\sim 6$. We compare further with Bridle \& King in Section 
\ref{Including Prior Information on Systematic Parameters}. It should be noted that the 
small effect of marginalising over the GI and II extra parameters 
may be a symptom of the parameterisation used. A full investigation of 
different GI and II paramerisations will be the subject of future investigations. 
\\
\\
\noindent{\bf c) Photometric Redshift Systematics, Intrinsic Alignments \& Shear Distortion}

It can be seen from Figure \ref{spec_bar_pz_phi_gi_ii} 
and from Table \ref{newline} that the effect of 
uncertainty in the shear distortion has a small effect however 
this could be due to the parameterisation used. In the case of the shear systematic terms, 
the values of $A_{\gamma}$, $\phi$ and $\gamma_{\rm bias}$ could be estimated from simulations 
(as is done in STEP; Heymans et al., 2006a; Massey et al., 2007) and 
the shear measurement method could be tuned (i.e. extra parameters added to minimise bias), 
or constructed \emph{ab initio} as in the case of 
Miller et al. (2007) and Kitching et al. (2008a), 
to minimise such effects (so that $A_{\gamma}\approx 1$, 
$\phi\approx 0$ and $\gamma_{\rm bias}\approx 0$). 

\begin{figure}
  \centering
 \psfig{figure=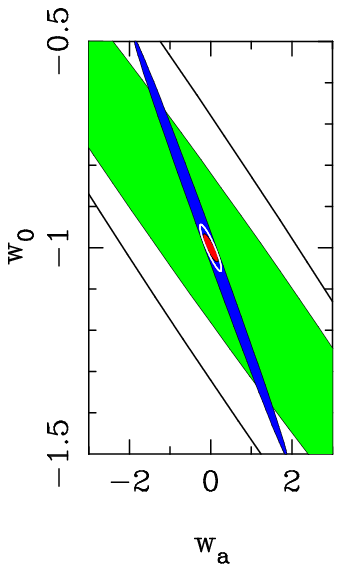,width=0.6\columnwidth,angle=0,clip=}
 \caption{The constraint from the shear-ratio method in the ($w_0$, $w_a$) plane. 
   The blue (darkest) solid ellipse is the Planck constraint, the green (lightest) 
   solid ellipse is the baseline 
   shear-ratio constraint alone, and the red (darker gray) 
   solid ellipse is the combined constraint without systematics. The 
   outer (black) solid line shows the shear-ratio constraint including systematic effects and
   the inner (white) solid line shows this constraint combined with the Planck prior. We stress
   again that this is the \emph{worst-case} degradation.}
 \label{geo_compare}
\end{figure}

For the shear-ratio method, including all three systematic effects and 
marginalising over all the free systematic parameters
(in this case only the extra photometric redshift parameters), the total reduction in the FoM is 
a factor of $2.3$ when combined with the Planck prior. Figure \ref{geo_compare} 
shows the total systematic effect in the ($w_0$, $w_a$) plane.
\begin{figure}
  \centering
 \psfig{figure=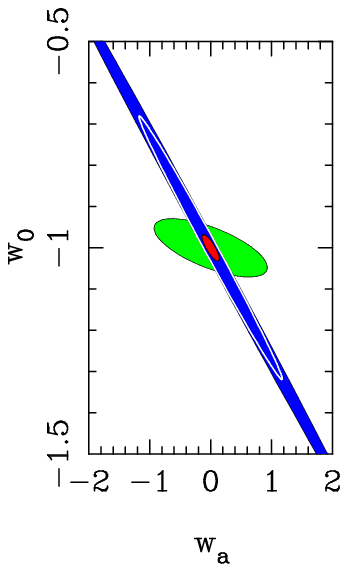,width=0.6\columnwidth,angle=0,clip=}
 \caption{The constraint from the 3D cosmic shear method in the ($w_0$, $w_a$) plane. 
   The blue (darkest) solid ellipse is the Planck constraint, the green (lightest) 
   solid ellipse is the baseline 
   3D cosmic shear constraint alone, and the red (darker gray) 
   solid ellipse is the combined constraint without systematics. 
   The inner (white) solid line shows the effect on the combined 3D cosmic shear and Planck prior 
   constraint including all systematic effect considered in this paper. We stress
   again that this is the \emph{worst-case} degradation.}
 \label{spec_compare}
\end{figure}

For 3D cosmic shear the effect of the photometric redshift systematics, 
shear distortion systematics and the intrinsic 
alignments in the ($w_0$, $w_a$) plane is shown in 
Figure \ref{spec_compare}. By including all three systematic effects, 
and marginalising over all free extra parameters, 
the FoM becomes FoM$=19$, a reduction in the FoM by a factor of $\sim 20$ 
relative to the baseline FoM. 

\subsection{Including Prior Information on Systematic Parameters}
\label{Including Prior Information on Systematic Parameters}
The marginalised results presented thus far have been in the self-calibration r\'egime, where 
the data itself is used to measure the extra systematic variables with no priors included. 
This presents a \emph{worst-case} 
scenario in the reduction of the FoM. In reality there should exist some information on systematic 
parameters either from different cosmological probes, simulations or data analysis techniques 
(such as 
photometric redshift code). Table \ref{priortab} shows the effect on the FoM 
of adding prior information on the systematic parameters when photometric redshift, 
intrinsic alignment 
and shear distortion systematics are included for both 3D weak lensing methods. 
We adopt a Gaussian prior of a given width $\sigma_P$ and use the 
same prior for all systematic parameters. 
It can be seen that in order to recover a substantial proportion 
($> 70\%$) of the baseline FoM a strong prior needs to be included with $\sigma_P=0.001$. 
The relative improvement between the two weak lensing methods is very similar. 

For simplicity we use a constant prior for all systematic parameters, 
in reality each systematic parameter will have a different prior. 
For example $z_{\rm bias}$ can currently be 
constrained to $\sim 2\%$. Abdalla et al. (2007) have demonstrated
that using neural net photometric redshift technique ({\tt AnnZ}) 
the bias can be constrained to $\pm 10^{-2}$.
This current level of constraint is still too large to effectively eliminate the FoM degradation, 
however the results from Abdalla et al. (2007) suggest that the IR bands will be able 
to improve the estimation of the bias significantly. Results from the Sloan Digital Sky Survey (SDSS; 
Adelman-McCarthy et al., 2005) shown that the photometric calibration can be known to $2\%$. 

STEP (Heymans et al, 2006a; Massey et al., 2007) 
has shown that shear measurement methods can be well calibrated
by simulations to within $\Delta A_{\gamma}\sim \pm 0.01$ and 
$\Delta \gamma_{\rm bias}\sim \pm 5\times 10^{-4}$. 
We have shown that marginalising over $\Delta A_{\gamma}$
causes a small change in the 
FoM so that dark energy constraints could be robust to distortions due to 
shear measurement given some improvement. However if a marginalisation is not done then the 
shear calibration needs to be biased by $\delta A_{\gamma} < 0.008$ (see Section 
\ref{Bias in Dark Energy Parameters}).
%The stability of shear measurement over size and magnitude is an unresolved issue and is 
%not investigated here.

Heymans et al. (2006) have probability distributions for the intrinsic alignment parameters used here 
with $\Delta A_{\rm II} \sim \pm 0.20\times 10^{-7}$ and 
$\Delta \theta_0 \sim \pm 0.4$ arcminutes. So that the 
prior errors here are overly optimistic for the amplitude of the intrinsic alignment 
terms and too optimistic 
for the scale dependence based on current simulations. However since the dependence of the 
FoM on these parameters is so small this should not affect our conclusions. 

\begin{table*}
\begin{center}
\begin{tabular}{|l|c|c|c|}
\hline
&Shear-Ratio FoM/FoM$_{\rm max}$&3D Cosmic Shear FoM/FoM$_{\rm max}$&Combined FoM/FoM$_{\rm max}$\\
\hline
Self Calibration&$0.43$&$0.11$&$0.83$\\
$\sigma_P=0.01$&$0.44$&$0.37$&$0.89$\\
$\sigma_P=0.001$&$0.60$&$0.90$&$0.92$\\
\hline
\end{tabular}
\caption{The effect on the predicted FoM of adding a Gaussian prior information on 
  the systematic parameters for the two 3D weak lensing methods. The self-calibration has no prior 
  information and is a worst case, $\sigma_P$ represents a constant 
  Gaussian prior of a given width on all the systematic parameters. FoM$_{\rm max}$ is the maximum 
  FoM given that all three systematic effects are included; for the shear-ratio method this is 
  FoM$_{\rm max}=360$ for 3D cosmic shear FoM$_{\rm max}=168$ and for the Combined constraint 
  FoM$_{\rm max}=594$. These are reduced from the absolute maximum, baseline FoM, by the 
  inclusion of the intrinsic alignment systematics.}
\label{priortab}
\end{center}
\end{table*}

\subsection{Prior on Photometric Redshifts}
It has been shown that the largest effect from a free systematic 
parameter on the FoM comes from the bias in the photometric redshifts 
for both 3D weak lensing methods. 
We should therefore expect that even 
adding prior information only on this parameter would improve the FoM.
\begin{figure}
 \psfig{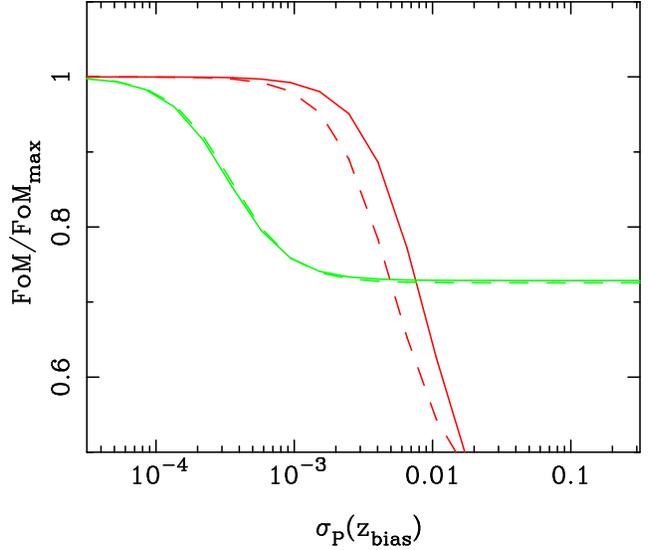}
 \caption{The change in the relative FoM with a Gaussian prior on $z_{\rm bias}$, assuming other 
   systematic parameters to be fixed for the 
   shear-ratio method (green/light, lines) and 3D cosmic shear (red/dark, lines). The solid lines 
   are without intrinsic alignment systematics and the dashed lines include intrinsic alignments.}
 \label{biasimprv}
\end{figure}
Figure \ref{biasimprv} shows the improvement in the relative FoM if a Gaussian prior is 
added to the $z_{\rm bias}$ constraint, marginalising over all other parameters which 
have no added prior. It can be seen that $\sim 80\%$ of the FoM  
can be recovered using a Gaussian prior with an error of $\sigma_P(z_{\rm bias})\sim 5\times 10^{-4}$ 
for the shear-ratio method and $\sigma_P(z_{\rm bias})\sim 2\times 10^{-3}$ for 3D cosmic shear when 
intrinsic alignment effects are included (dashed lines). 

To begin to improve upon the FoM for which no prior has been added the shear-ratio method requires a 
prior with an error that is approximately $10$ times smaller than for the 3D cosmic shear method. 
This factor of $10$ between the two methods 
reflects the results of Section \ref{Bias in Dark Energy Parameters} 
and Heavens et al. (2006), the 3D cosmic shear method is less 
sensitive to this parameter so that in the self-calibration r\'egime, with a poor prior, 
the reduction in the FoM is larger. 
Since the constraint on $z_{\rm bias}$ is already smaller for the shear-ratio method, as can be seen 
from Figure \ref{w0zb}, the 
extra prior error needs to be even smaller to begin to improve upon the constraint. 

The solid lines in Figure \ref{biasimprv} show the improvement in the FoM as the prior on 
$z_{\rm bias}$ is improved where intrinsic alignment effects have not been included. For the 
shear-ratio method this has little effect, since the method is relatively insensitive to intrinsic 
alignment effects as discussed in Section \ref{Discussion}. 
For 3D cosmic shear the requirement on the accuracy of 
$z_{\rm bias}$ to recover $\sim 80\%$ of the FoM 
is relaxed to $\sigma_P(z_{\rm bias})\sim 6\times 10^{-3}$, a factor of $\sim 3$ times larger 
than when intrinsic alignments are included. This 
is in agreement with the results of Bridle \& King (2007) who find that the average 
photometric redshift
error needs to be $\sim 3$ to $4$ times smaller to recover $\sim 80\%$ of the FoM when intrinsic 
alignements are included.

Comparing the two 3D shear methods, we find that
the 3D cosmic shear has slightly better ideal FoM, but this degrades more if
the systematics need to be estimated from the data themselves.  The better
the prior is on the systematics, the better the 3D cosmic shear method will fare, but
with reasonable priors which should be achievable with current techniques,
the shear-ratio method and 3D cosmic shear should attain comparable accuracy.

\subsection{Spectroscopic Calibration of Photometric Redshifts}

Spectroscopic redshifts could be used to calibrate the redshift bias. Assuming Poisson statistics
the number of spectroscopic redshifts required can be written as 
\be 
\label{s1}
\sigma(z_{\rm bias})=\frac{\sigma_z(z)}{\sqrt{N_{\rm spec}}}.
\ee
Using the result from Taylor et al. (2007) this can also be related to the bias parameter,  
defined in equation (\ref{biaseq}), by
\be
\label{s2}
N_{\rm spec}=\left[\frac{C_{\rm bias}\sigma_z(z)}{\delta w(z_p)}\right]^2.
\ee
For the shear-ratio method Figure \ref{biasimprv} shows that a prior error of 
$\sigma_P(z_{\rm bias})\sim 5\times 10^{-4}$ is needed to eliminate the effect of marginalising 
over $z_{\rm bias}$. Assuming an average redshift error of $\sigma_z(z)\sim 0.025$ 
and using equation (\ref{s1}), this implies $N_{\rm spec}\sim 3\times 10^{3}$. 
If $z_{\rm bias}$ is assumed to be fixed but biased then 
from Table \ref{biastable} we have $C_{\rm bias}\sim 55$ for the shear-ratio method. If the bias in 
$w(z_p)$ is required to be less than $\delta w(z_p)\sim 0.01$ then 
equation (\ref{s2}) implies that $N_{\rm spec}\sim 2\times 10^{4}$. 
These predicted numbers of spectroscopic redshifts is in agreement 
with predictions for weak lensing tomography, for example Abdalla et al. (2007).

For 3D cosmic shear Figure \ref{biasimprv} shows that a prior error of 
$\sigma_P(z_{\rm bias})\sim 2\times 10^{-3}$ is required to recover 
the original FoM, using equation (\ref{s1}) this implies the 
number of spectroscopic redshifts needed to 
be $N_{\rm spec}\sim 6\times 10^{2}$. Assuming $z_{\rm bias}$ is fixed it can be deduced from Table 
\ref{biastable} that $C_{\rm bias}\sim 5.48$ which implies, using equation (\ref{s2}) that 
$N_{\rm spec}\sim 2\times 10^{2}$. This relatively small calibrating sample is in 
approximate agreement with the results of Heavens et al. (2006). This is as a result of the small 
degeneracy between $z_{\rm bias}$ and $w_0$ as discussed in Section \ref{Discussion}.
\begin{figure}
 \psfig{figure=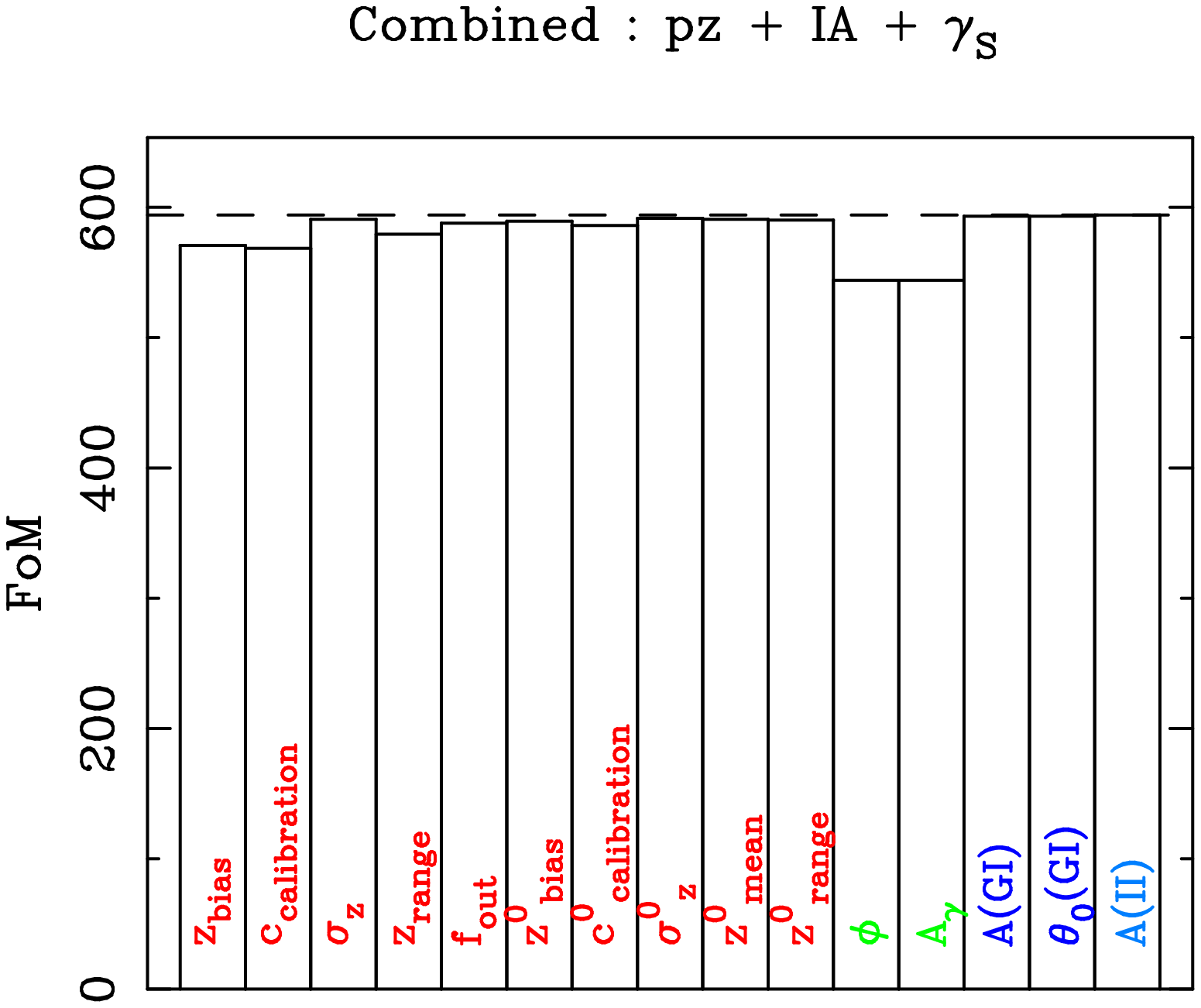,width=\columnwidth,angle=0}
 \caption{The change in the figure of merit with each of the individual 
   systematic parameters combining the shear-ratio and 3D cosmic shear methods, 
   including photometric redshift, shear distortion, GI and II systematic effects. 
   The dashed line shows the baseline combined FoM.}
 \label{combined_bar}
\end{figure}

\subsection{Combined constraints \& Bottom-Line Predictions}
\label{Combined constraints}
By combining the constraints from the shear-ratio and 3D cosmic shear method the relative 
decrease in the FoM may be smaller due to parameter degeneracies between the systematic parameters 
being lifted, and the baseline FoM will be larger as the dark energy constraints are combined. 

The combination presented here does not take into account the full covariance between the two 
methods but should be valid to a first approximation since the shear-ratio uses clusters (small 
scale features) and the majority of the dark energy signal for the 
3D cosmic shear method is from approximately sub-degree scales 
(the maximum signal is at $\ell \approx 1000$; Heavens et al., 2006). 
Also the full correlation should only appear between the noise 
of the shear-ratio method and the signal of 3D cosmic shear both of which depend on the matter
power spectrum. 
\begin{figure}
 \psfig{figure=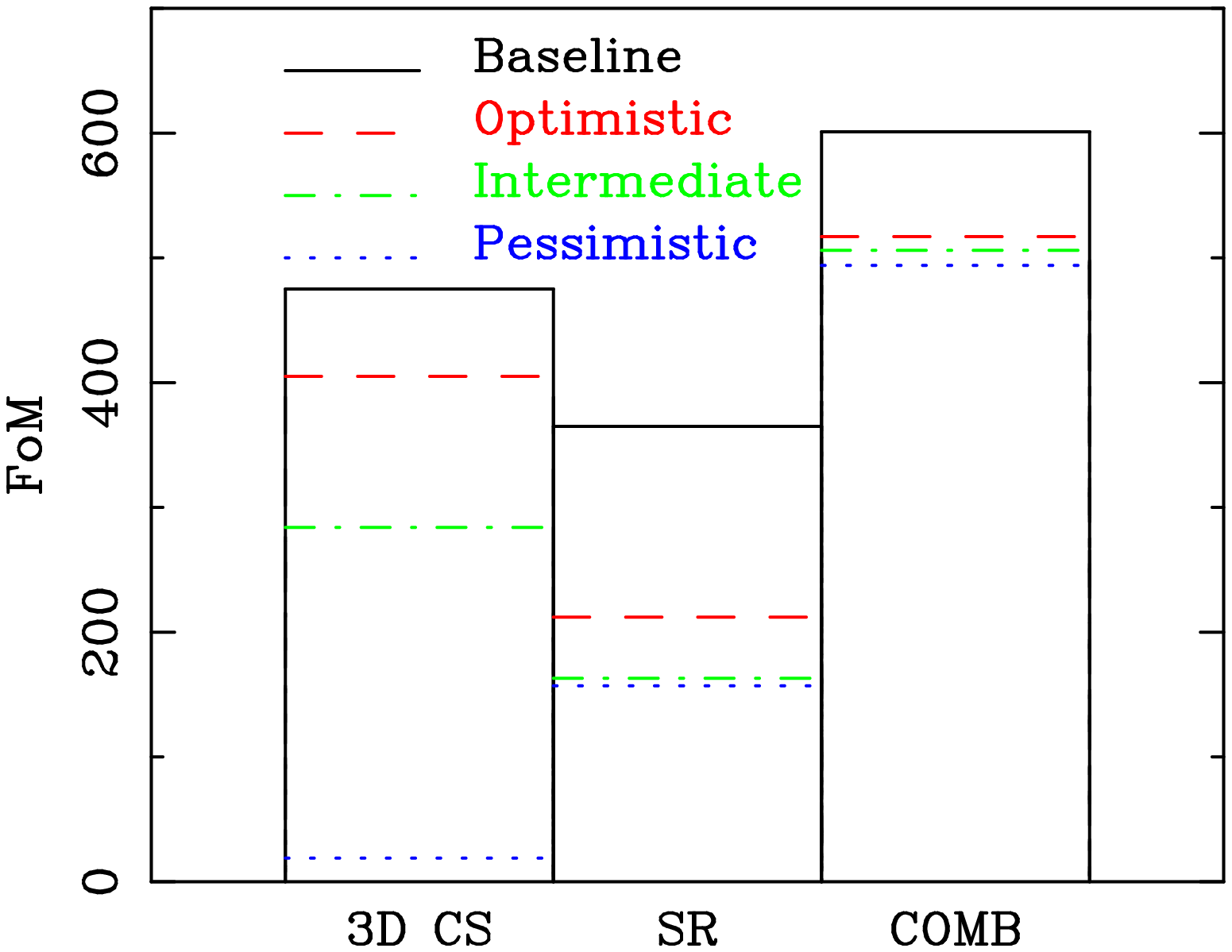,width=\columnwidth,angle=0}
 \caption{The change in the FoM for a DUNE-type survey using
   3D cosmic shear (3D CS), the shear-ratio method (SR) 
   and the combined FoM (COMB) for each of the optimistic, intermediate and pessimistic scenarios. 
   Here the absolute FoM is shown for the fiducial survey considered.}
 \label{bottom_bar}
\end{figure}
\begin{figure}
 \psfig{figure=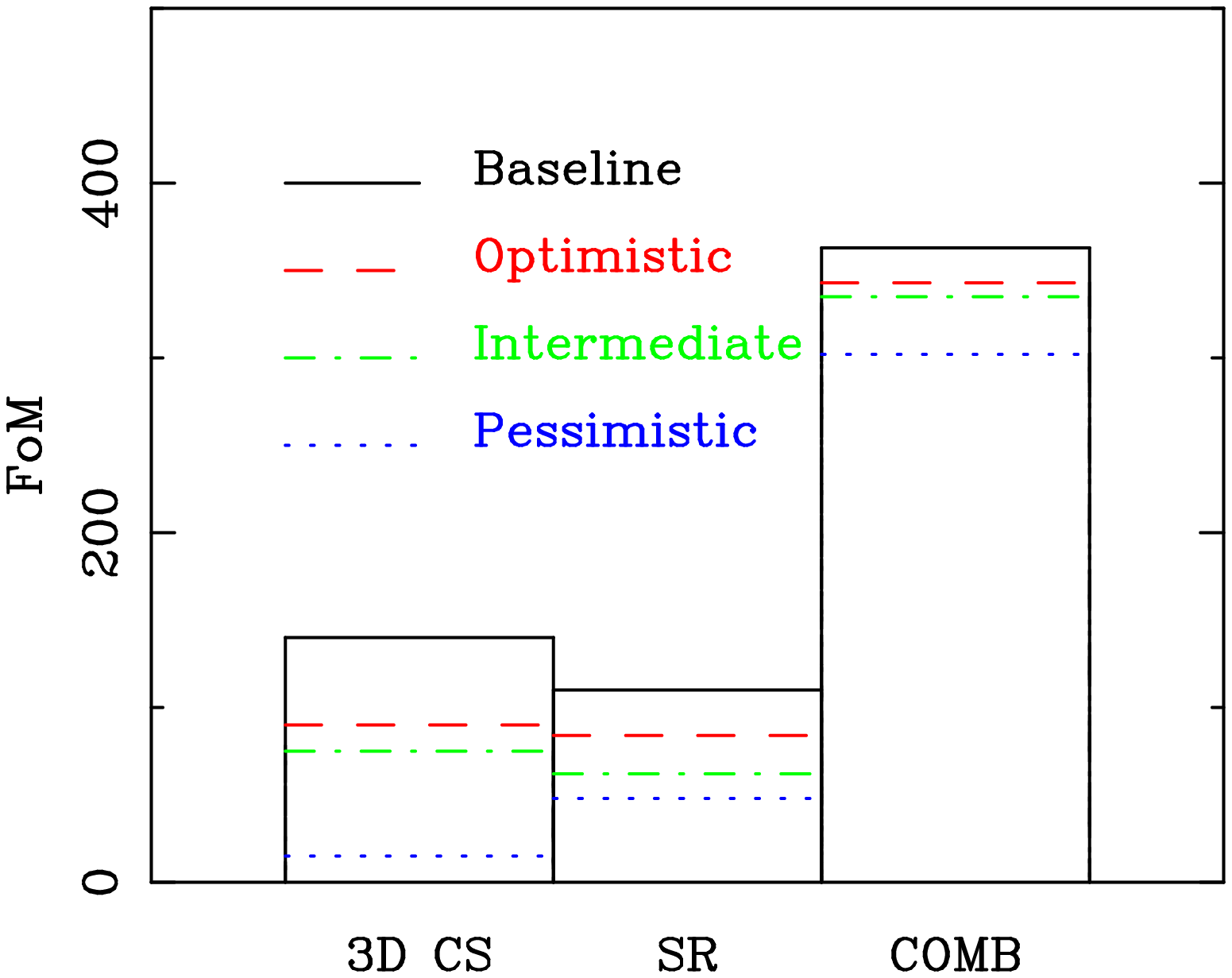,width=\columnwidth,angle=0}
 \caption{The change in the FoM for a PS1-type survey using 
   3D cosmic shear (3D CS), the shear-ratio method (SR) 
   and the combined FoM (COMB) for each of the optimistic, intermediate and pessimistic scenarios. 
   Here the absolute FoM is shown for the fiducial survey considered.}
 \label{bottom_bar_ps1}
\end{figure}

The baseline FoM for the fiducial survey when the methods are combined, with a Planck prior included,
is FoM$=601$ with pivot redshift error of $\Delta w(z_p)=0.015$. 
Figure \ref{combined_bar} shows the effect of the combined FoM on the systematic parameters. It can 
immediately be seen that the combined FoM is substantially larger than either method alone, and that
the degradation of the FoM with the systematic parameters is much smaller. In particular the 
methods' different parameter degeneracies are very complementary for $z_{\rm bias}$ and 
$c_{\rm cal}$. Since the shear-ratio method provides no constraint on the shear distortion 
parameters these have a relatively large effect on the combined FoM via the 
3D cosmic shear method's dependence. 
When marginalising over all systematic parameters the FoM$=494$ with a pivot redshift error 
of $\Delta w(z_p)=0.018$ with a Planck prior included.  

This is an encouraging result, despite the naive addition of the two methods, it appears that the 
3D cosmic shear and shear-ratio methods are very complementary in terms of both cosmological and 
systematic parameter degeneracies. 

In Table \ref{bottomline} and Figures \ref{bottom_bar} and \ref{bottom_bar_ps1} 
we present the effect on the relative and absolute FoM given 
three different future scenarios, now showing results for both a Pan-STARRS survey and the 
DUNE-type fiducial survey. 

Pessimistically one could 
assume that the II and GI effects cannot be removed from data and that there are no prior 
constraints available for photometric redshift parameters. 
This worst-case seems unlikely since already
using current photometric redshift codes one could provide priors for the bias and calibration of 
a photometric redshift sample (e.g Abdalla et al, 2007) 
and there are planned and ongoing spectroscopic 
instruments that could provide adequate calibration. This result is a worst-case scenario, 
since it has been shown 
(for example Heymans \& Heavens, 2003;  Heymans et al., 2004; King \& Schneider, 2003; King, 2005)
that intrinsic alignments can be removed to some extent from cosmic shear data. The worst-case 
FoM$\sim 20$ for the 3D cosmic shear is approximately the same for both surveys so 
that the relative change for DUNE is larger, in this r\'egime so much information 
is lost through systematic effects that the differences in survey design have a sub-dominant effect. 

The intermediate stage 
takes into account current belief in the likely removal 
of the II intrinsic alignment term and includes a 
currently realistic prior on the photometric redshift systematic parameters; 
this is currently the most realistic scenario. There is also broad agreement between the relative 
FoM changes given the different survey designs. We conclude that such a 
scenario could result in a factor of $\sim 2$ reduction in the 
statistical FoM for both 3D weak lensing 
methods. 

Going beyond current photometric redshift constraints and understanding of intrinsic alignments 
we present an optimistic conclusion based on the assumptions that both GI and II effect could be 
removed and the photometric redshift systematic parameters could be calibrated to $1\times 10^{-3}$.
Again there is broad agreement over survey design. In this case the 3D cosmic shear constraints 
are degraded by $\sim 1.2$ and 
the shear-ratio constraints show a FoM reduction by a factor of $\sim 1.7$ for the fiducial survey.
By combining the 3D cosmic shear and shear-ratio constraints parameter degeneracies are 
lifted to such an extent that even in the pessimistic case the reduction 
in the FoM due to all three primary systematics effects is only a factor of $1.2$. 

Given the agreement between two different 3D weak lensing methods and over two different survey 
designs we conclude that one may expect \emph{at most} 
a factor of $2$ reduction in the FoM using 3D 
weak lensing as a result of the three primary systematic effect considered in this paper. We expect that this factor would be substantially less in actuality given that shear-galaxy correlations could be used and the weak lensing methods can be combined as shown.   

\begin{table*}
\begin{center}
\begin{tabular}{|l|l|c|c|c|c|c|}
\hline
Survey&&IA&Photo-z Prior&SR FoM/FoM$_{\rm max}$&3D CS FoM/FoM$_{\rm max}$&Comb FoM/FoM$_{\rm max}$\\
\hline
DUNE (fiducial)&Optimistic& None & $\sigma_P=0.001$     &$0.58$  &$0.85$ & $0.86$\\
PS1 &Optimistic& None & $\sigma_P=0.001$     &$0.76$  &$0.65$ & $0.94$\\
\hline
DUNE (fiducial)&Intermediate& GI only & $\sigma_P=0.01$ &$0.45$  &$0.59$ & $0.84$\\
PS1 &Intermediate& GI only & $\sigma_P=0.01$ &$0.56$  &$0.54$ & $0.92$\\
\hline
DUNE (fiducial)&Pessimistic&II and GI & None            & $0.43$ &$0.04$ &$0.82$\\
PS1 &Pessimistic&II and GI & None            & $0.44$ &$0.11$ &$0.83$\\
\hline
\end{tabular}
\caption{Here we present three possible scenarios representing optimistic, intermediate 
  and pessimistic points of view on the future development of weak lensing systematics using 
  3D cosmic shear (3D CS), the shear-ratio method (SR) and the combining the methods (COMB)
  The pessimistic 
  scenario is unlikely given that II effects can be removed from data and the current photometric 
  redshifts have at least some prior on their bias and calibration. Note here FoM$_{\rm max}$ is the 
  baseline FoM given in Table \ref{baseline}, not the reduced maximum FoM 
  used in Table \ref{priortab} which already included in intrinsic alignment systematics.}
\label{bottomline}
\end{center}
\end{table*}

\section{Conclusion}
\label{Conclusion}

We have shown using simple analytic approximations that systematic effects can have a substantial
impact on the ability of 3D weak lensing methods, the shear-ratio method and the 3D 
cosmic shear method, to constrain the dark energy equation of state.
We used the Figure of Merit (FoM) to gauge the ability of a next generation experiment 
to constrain the dark energy equation of state. The systematic effects we 
considered are those associated with photometric redshifts, an overall distortion in the image 
plane and intrinsic alignments (both the GI and II terms). 

The dark energy FoM can be degraded a factor of $2$ due to the photometric 
redshift systematics alone, for both 3D weak lensing methods. 
Shear distortion systematics have a small 
effect for both methods. Intrinsic alignment effects alone can degrade the FoM by a further 
factor of $2$ for the 3D cosmic shear method, but have a small effect for the shear-ratio method. 
This difference is due to the way in which the methods use shear information. 

%The majority of this paper has presented results in which the extra systematic parameters have no prior
%information on them, so that they are assumed to be constrained from the weak lensing data itself -- 
%the so called self calibration r\'egime. The results which do not include any prior information 
%are therefore a worst-case scenario, in reality we would expect there to be some 
%prior information on all of the systematic parameters. 

When an extra systematic parameter is encountered it can either be marginalised over using the 
available data (equivalent to self-calibration), 
thereby increasing the marginal error on any cosmological parameter of interest, 
or it can be assumed to be fixed. If a parameter is fixed then any deviation away from the 
assumed value will bias the most likely value of any measured cosmological parameter. This 
bias is a function of both the cosmological parameter error and the sensitivity of a method to 
any systematic parameter. From this analysis it has been shown that assuming some parameters 
to be fixed can lead to large biases in $w(z_p)$ and $w_a$, this is complimentary to the analysis 
done by Amara \& Refregier (2007) 
who investiagted the bias in cosmological parameters due to shear 
measurement systematics using weak lensing tomography.

The methods are remarkably insensitive to many parameters, in particular most of the photometric
redshift parameters, including the fraction of outliers, 
and an overall distortion of the shear field. 
By adopting a parameterisation of the photometric-spectroscopic plane we have shown that 
a bias in any photometric redshift redshift technique needs to be known to within 
$\pm 10^{-3}$ for the dark energy FoM to remain unaffected. 
We have shown that 
to calibrate the photometric redshift parameters approximately $10^5$ 
spectroscopic redshifts are required for the shear-ratio method and approximately
$10^2$ to $10^3$ for the 3D cosmic shear method. 
This difference can be attributed to the binning in 
redshift required by the shear-ratio method, which may also explain the agreement between the 
predicted spectroscopic requirements of the shear-ratio and shear tomography methods. 

The intrinsic alignment terms were modelled using the Heymans et al. (2006) 
analytic approximations. The GI and II terms had a small effect on the FoM 
from the shear-ratio method,
we found a drop in the FoM of approximately $10\%$ when the extra covariances were included. For 
3D cosmic shear the FoM is reduced by approximately $50\%$, but there is a very small sensitivity 
to the extra intrinsic alignment systematic parameters. 
The 3D cosmic shear result is in agreement to what has been found using shear tomography 
in Bridle \& King (2007). The caveat to these comparisons is that we use a fully 3D cosmic shear 
method, with no binning, and only investigate 
the Heymans et al. (2006) parameterisation; Bridle \& King (2007) consider a variety of 
parameterisations and investigate a binning tomographic method. 

We have shown that a good prior on systematic parameters can improve the FoM. The relative reduction 
in the FoM can be limited to $\leq 30\%$ if the prior on all systematic parameters 
has a Gaussian error 
of $\sigma_P=0.001$. In particular we have shown that a prior on $z_{\rm bias}$ can improve the FoM. 
Good priors on $z_{\rm bias}$, $c_{\rm calibration}$ would be particularly helpful in limiting the 
effect on the FoM.

By combining the 3D weak lensing methods the FoM can be increased by a up to a factor of $6$ 
relative to the methods individually. Furthermore the photometric redshift 
systematic parameter degeneracies are complementary leading to less systematic degradation in the 
combined constraints.  

%We have highlighted two types of systematic effect. Those that appear in the signal part of a 
%method are included by marginalising over any extra systematic parameters. These systematic effects 
%can be removed/improved upon by increasing the prior knowledge of any extra parameters. The 
%second type of effect are those that enter as an extra covariance (signal or noise terms) 
%in any method and cannot be completely removed by simply including prior information.

%Photometric redshift, shear distortion and intrinsic alignment effects can have
%a dramatic, but potentially controllable, impact on the dark energy parameter estimation 
%using data from future weak lensing 
%surveys. We conclude that given realistic assumptions on the future development in understanding of the 
%three primary systematic effects the FoM from 3D weak lensing will be reduced by at most a factor of $2$
%due to these systematic effects. 
%With current constraints from photometric redshift estimation and shear measurement methods 
%the FoM will be degraded by a factor of $2$ from the ideal case due to this systematic effects. 
%However given modest improvements in photometric redshift estimation, 
%accurate shear measurement methods and an 
%understanding of intrinsic alignment effects such systematics have the potential to be 
%alleviated. 
%Finally it should be emphasised that the analytic approach of this paper to systematic effects 
%should be superceded by full (n-body) ray tracing simulations of future surveys.

The bottom line is that the most important systematics to control are those
concerning the photometric redshift distribution.  If these can be
controlled to 1\% accuracy, then the FoM for proposed future surveys such as
DUNE and Pan-STARRS may be reduced by at most a factor of order two. 
In order to reduce these systematics 
to a negligible level needs $0.001$ accuracy in median redshifts, 
requiring ${\mathcal O} (10^4)$ redshifts. We make a number of recommedations and observations 
with which to guide future systematic investigations
\begin{itemize}
\item
Photometric redshift systematics play a dominant role in the systematics 
that affect 3D weak lensing. 
The individual systematic parameter which can have the largest effect on the FoM is the 
bias in photometric redshifts. However 
approximately ${\mathcal O} (10^4)$ spectroscopic redshifts should be enough to calibrate 
photometric redshifts to the required accuracy. These would need to be representative 
of the photometric galaxies and complete.   
\item 
If shear calibration bias is assumed to be fixed then an uncertainty in the bias of $\sim 0.008$
can bias dark energy parameters by $>0.01$. However marginalising over shear bias has a smaller 
effect on the FoM. 
\item
Intrinsic alignments play a major role in 3D weak lensing systematic effects, and can reduce the 
maximum achievable FoM by up to $4$. The broad agreement between the parameterisations investigated 
here and in Bridle \& King (2007) suggest that the general trends are robust. 
\end{itemize}

Despite degrading systematic effects   
3D weak lensing retains the potential to be the most powerful cosmological probe of dark energy. 
Weak lensing is entering a formative period in its development, given that the statistical ability of 
the method to constrain cosmology is accepted attention must now be focussed on 
understanding and reducing systematic effects.

\section*{Acknowledgments}
TDK is supported by the Science and Technology Facilities Council,
research grant number E001114. This work was partly supported by the DUEL EC RTN Network 
(contract number MRTN-CT-2006-036133). 
We thank Filipe Abdalla, Adam Amara, Sarah Bridle, Catherine Heymans, Lance Miller, 
Alexandre Refregier and Chris Wolf for insightful discussions.

\newpage
\section*{Appendix A}

In this Appendix we will present the technical details of how the primary systematic effects are 
included in the shear-ratio method. For full details on the theory and implementation of the method 
see Taylor et al. (2007) and Kitching et al. (2007). 

\subsection*{Shear-ratio photometric redshift systematics}
\label{Geometric shear-ratio photometric redshift systematic}
The shear-ratio method takes the ratio of average 
tangential shear in different redshift bins behind a cluster. 
The binning is necessary since if it is assumed that observed ellipticities have zero mean and that 
the distribution of intrinsic ellipticities is Gaussian the resulting 
distribution of the ratio of 
ellipticities has a Cauchy/Lorentzian distribution (and hence infinite variance). 
Around a cluster the 
mean ellipticity is non-zero and so Gaussian errors can be assumed, and this is certainly so when 
binning around a cluster and in redshift, so that the number of galaxies contributing to the mean is 
increased. 

Photometric redshift systematics affects the shear-ratio 
signal via the uncertainty in a galaxies position meaning 
it could be scattered out of (or in to) a bin i.e. the galaxies used to infer the tangential shear in 
any given bin could in actuality be in another bin. The average tangential shear in the $i^{\rm th}$ 
redshift bin behind a cluster at redshift $z_l$ is given by 
\ba
\label{A1}
\lgl \gamma_{t,i} \rgl = \gamma_{t,\infty}
&&\int_{z_l}^\infty \!\! dz \, \frac{S_k[r(z)-r(z_l)]}{S_k[r(z)]}
n(z)W(z)\nn
&&\int_{z_i-\Delta z/2}^{z_i+\Delta z/2}
\!\! dz' \, p(z-z'|\sigma_z),
\ea
where $S_{[k=-1,0,+1]}(r)=[\sinh(r),r,\sin(r)]$, $\Delta z$ is the width of the bin, $n(z)$ is the 
number density of source galaxies and 
$\gamma_{t,\infty}$ is the tangential shear for a source galaxy at $z=\infty$, which is cancelled 
in taking the ratio of shears. 

$p(z-z'|\sigma_z)$ is the probability distribution for a galaxies position in redshift. 
If $p(z-z'|\sigma_z)$ is Gaussian then equation (\ref{A1}) simplifies to 
(Taylor et al., 2007; equation 25)
\be
\frac{\lgl \gamma_{t,i}\rgl}{\gamma_{t,\infty}}=
\int_{z_l}^\infty \!\! dz \, n(z)
\frac{S_k[r(z)-r(z_l)]}{S_k[r(z)]}
P_{\Delta z}[z_i-z |\sigma_z(z_i)]W(z).
\label{mean_gamma}
\ee
For a sum of Gaussians each integral is simplified in the same way. 
The integrand $P_{\Delta z}[z_i-z |\sigma_z(z_i)]$ is given by (e.g., Ma et al., 2005)
\ba
P_{\Delta z}[z |\sigma_z]=\frac{1}{2}\left[{\rm erf}
\left(\frac{z+\Delta z/2}{\sqrt{2}\sigma_z}\right)\right]\nonumber\\
-\frac{1}{2}\left[{\rm erf}\left(\frac{z-\Delta z/2}{\sqrt{2}\sigma_z}\right)\right]
\ea
which is the part of the redshift error distribution that lies in a
redshift bin centred on $z$ of width $\Delta z$, and ${\rm
  erf}(x)$ is the error function. The function $W(z)$ in equation (\ref{mean_gamma}) is given by
\be
\label{GEOeq25a}
W(z)=\frac{\widetilde w(z)}{\int_0^{\infty}dz'\widetilde w(z')n(z')P_{\Delta
    z}[z-z' |\sigma_z(z)]}
\ee
where $\tilde w(z)$ is some arbitrary weighting function of the shears
in redshift, which we take as $\tilde w(z)=1$.

So any extra parameters used to describe the photometric redshift probability distribution 
enter the signal of shear-ratio method via the determination of the mean tangential shear in a 
given redshift bin. 

\subsection*{Shear-ratio intrinsic alignments}
\label{Geometric shear-ratio intrinsic alignments}
Recasting the measured average tangential shear with an extra intrinsic shear component, 
$\gamma_i\rightarrow\gamma_i+e_i$ where $e_i \ll \gamma_i$ the shear-ratio becomes
\be
\label{e3}
\frac{\gamma_i'}{\gamma_j'}=\frac{\gamma_i+e_i}{\gamma_j+e_j}\simeq\frac{\gamma_i}{\gamma_j}
\left( 1+\frac{e_i}{\gamma_i}-\frac{e_j}{\gamma_j}\right)
\ee
where the shear is the average in an aperture about a cluster in a given redshift bin. 
Since it is assumed that any intrinsic effect is small the shear-ratio signal 
is unaffected to first order by intrinsic alignment effects. 
The noise covariance properties of 
the shear ratio are affected in the following way. Taking the 
covariance of the shear ratio for two different pairs of bins ($i,j$) and ($m,n$) 
where $z_i<z_j$ and $z_m<z_n$ gives
\ba
\label{e4}
\left\langle\frac{\delta R_{ij}}{R_{ij}}
\frac{\delta R_{mn}}{R_{mn}}\right\rangle&=&
\frac{C^{\gamma \gamma}_{1,im}}{\gamma_{i} \gamma_{m}}+
\frac{C^{\gamma \gamma}_{1,jn}}{\gamma_{j} \gamma_{n}}\nn
&-&\frac{C^{\gamma \gamma}_{1,in}}{\gamma_{i} \gamma_{n}}-
\frac{C^{\gamma \gamma}_{1,jm}}{\gamma_{j} \gamma_{m}}\nn
&+&\frac{C^{\gamma \gamma}_{2,i,{\rm min}(j,n)}}{\gamma^2_{j}}\delta^K_{im}+
\frac{C^{\gamma \gamma}_{2,{\rm max}(i,m),j}}{\gamma^2_{j}}\delta^K_{jn}\nn
&+&\frac{C^{\gamma e}_{im,i>m}}{\gamma_i\gamma_m}+
\frac{C^{\gamma e}_{jn,j>n}}{\gamma_j\gamma_n}\nn
&+&\frac{C^{\gamma e}_{mi,m>i}}{\gamma_i\gamma_m}+
\frac{C^{\gamma e}_{nj,n>j}}{\gamma_i\gamma_m}\nn
&-&\frac{C^{\gamma e}_{jm}}{\gamma_j\gamma_m}-
\frac{C^{\gamma e}_{mj}}{\gamma_m\gamma_j}-
\frac{C^{\gamma e}_{in}}{\gamma_i\gamma_n}-
\frac{C^{\gamma e}_{ni}}{\gamma_n\gamma_i}\nn
&+&\frac{C^{e e}_{im}}{\gamma_{i} \gamma_{m}}-
\frac{C^{e e}_{in}}{\gamma_{i} \gamma_{n}}\nn
&+&\frac{C^{e e}_{jn}}{\gamma_{j} \gamma_{n}}-
\frac{C^{e e}_{jm}}{\gamma_{j} \gamma_{m}}.
\ea  
The first six terms are due to sample shear covariance, described in Taylor et al. (2007). 
The next four terms are the correlations between the intrinsic ellipticity of one bin with 
the tangential shear in another bin, correlations between either foreground bins or either 
background bins. Note that $e_m$ is only correlated with $\gamma_i$ 
if $z_i>z_m$ i.e. $C^{\gamma e}_{im,i>m}=\langle d\gamma_i de_m \rangle$ for all $z_i>z_m$.
The next four terms, which have a negative constribution, are the correlations between the 
intrinsic ellipticity of a foreground bin with the shear of 
a background bin $C^{\gamma e}_{jm}=\langle d\gamma_j de_m \rangle$. The last four terms are the 
correlations between the intrinsic ellipticities between bins.   
Equation (\ref{e4}) is as expected since 
there are no correlations between the shear in a given bin and the 
intrinsic ellipticity in that bin i.e. $\langle d\gamma_i de_i \rangle=0$. 

Since we use the tangential shear in an aperture about a cluster, equation (\ref{e2}) 
needs to be averaged. Assuming an aperture of size $\theta$
\ba
\label{e5}
C^{\gamma e}_{ij}=\langle \gamma_i e^*_j\rangle=&&\frac{2A_{\rm GI}E(z_i,z_j)}{\pi\theta^4}\nn
&&\int_0^{\theta}d\tilde{\theta}\tilde{\theta}
\int_0^{\theta}d\theta'\theta'\nn
&&\int_0^{2\pi}d\phi
\frac{1}{\sqrt{\tilde{\theta}^2+\theta'^2-
    2\tilde{\theta}\theta'\cos(\phi)}+\theta_0}
\ea
which can be calculated numerically, where $E(z_i,z_j)=D(z_j)D(z_i-z_j)/D(z_i)$. We use values of  
$A_{\rm GI}=-0.92\times 10^{-8}$ $h {\rm Mpc^{-1}}$ arcmin and $\theta_0=1.32$ arcmin 
for tangential shear taken from Heymans et al. (2006). The II terms are given by
\be
C^{e e}_{ij}=\frac{A_{\rm II}}{1+(r[z_j]-r[z_i]/B_{\rm II})^2}
\ee
where $A_{\rm II}=0.45\times 10^{-3}$ and $B_{\rm II}=1 h^{-1}$Mpc. 
$C^{e e}_{ij}$ should be very small since $z_i \not= z_j$ and the bin widths used, as a result of the 
photometric redshift error $\sigma_z(z)=0.025(1+z)$, correspond to 
$\Delta r(z)=r[z_j]-r[z_i]\gg B_{\rm II}$. 

Note that since the intrinsic alignment effect does not enter into the signal of the geometric 
shear-ratio method the parameters $A_{\rm GI}$ and $\theta_0$ cannot be 
marginalised over but rather a further noise term is added to the covariance. 

\subsection*{Shear-ratio image distortion}
\label{Geometric shear-ratio image distortion}
Since we assume a Singular Isothermal Sphere (SIS) profile (see Taylor et al., 2007) the shear 
can be written
\ba
\label{e8}
\gamma&=&\left[\frac{\theta_0}{2|\theta|}A_{\gamma}{\rm e}^{-i2\phi}
+\gamma_{\rm bias}\right]{\rm e}^{i2\phi_C}\nn
\gamma_t&=&-\Re[\gamma{\rm e}^{-2i\phi_C}]=-\frac{\theta_0}{2|\theta|}A_{\gamma}\cos(2\phi)
-\gamma_{\rm bias}
\ea
where $\phi_C$ is the azimuthal position of the a galaxy relative to the 
cluster center and $\phi$ is the 
unknown systematic rotation of a galaxy's shape relative to the center of the galaxy's image. 
The shear-ratio signal takes the ratio of tangential shear so 
any overall distortion of the shear will cancel to first order, 
so that the signal remains unaffected. However the noise 
properties are affected since the 
cross-component shear, which can be used to estimate the scatter in $\gamma_t$
(Kitching et al., 2007) becomes 
\be
\label{e9}
\langle\gamma_{\times}\rangle=\left(\frac{\theta_0}{2|\theta|}\right)A_{\gamma}\sin(2\phi)+
\gamma_{\rm bias}.
\ee
Using the notation of Taylor et al. (2007) the fractional error in the tangential 
shear due to the extended description of the shear is, to first order in $\gamma_{\rm bias}$, 
\ba
\label{e10}
\left(\frac{\Delta\gamma_i}{\gamma_i}\right)_{\rm sys}^2
&=&\frac{\tan^2(2\phi)}{2\pi n_0 \Delta z_i\theta^2}\nn
&+&\frac{\tan^2(2\phi)\gamma_{\rm bias}}{\pi n_0\Delta z_i\theta_0 A_{\gamma}\theta}
\left(\frac{1}{\sin(2\phi)}-\frac{1}{\cos(2\phi)}\right)\nn
&-&\frac{\sigma^2_{\epsilon}\gamma_{\rm bias}\theta}{\pi n_0\Delta z_i A^3_{\gamma}\theta^3_0\cos^3(2\phi)}
\ea
so that the overall fractional error in the shear is now
\be 
\label{e11}
\left(\frac{\Delta\gamma_i}{\gamma_i}\right)^2=
\frac{\sigma^2_{\epsilon}}{2\pi n_0 \Delta z_i\theta_0^2 A_{\gamma}\cos(2\phi)}+
\left(\frac{\Delta\gamma_i}{\gamma_i}\right)_{\rm sys}^2.
\ee

The effect of such the extra systematic terms is to increase the noise in the 
shear-ratio method. A caveat to this simple parameterisation is that any 
locally non-uniform distortion, for example if $A_{\gamma}\rightarrow A_{\gamma}(z)$ or 
$\phi\rightarrow \phi(\theta)$, 
would not cancel in taking the ratio and any
parameters describing such effects as a function of redshift or position 
would have to be marginalised over. 

\section*{Appendix B}
This Appendix will present in technical detail the effect of each primary systematic parameterisation 
on the 3D cosmic shear method. For a full exposition of the method see Castro et al. (2005), 
Heavens et al. (2006) and Kitching et al. (2007).

We also take this oppurtunity to highlight an \emph{erratum} present in Heavens et al. (2006) and 
Kitching et al. (2007). The signal covariance prefactor should not have a factor of $H_0^4$ if 
$k$ has units of Mpc$^{-1}$. If $k$ has units of $h$Mpc$^{-1}$ then one should be present. This 
can be seen straightforwardly from Poisson's equation which enters the shear transform (equations 
17 to 21; Heavens et al., 2006)
\be
\Phi(k,l)=-\frac{3 \Omega_m H_0^2}{2 k^2 a}\delta(k,l;r).
\ee
If $k$ is in $h$Mpc$^{-1}$ then this should give no $h$ dependence. Because Heavens et al. (2006) 
were using $k$ in units of Mpc$^{-1}$ this resulted in the transform being proportional to $h^2$ 
so that the covariance was proportional to $h^4$. 

Also in calculating the redshifts these papers used: $z_i=z_{i-1}+(dz/dr)dr$ 
where $dz/dr \propto h$, but $dr$ is $\propto h^{-1}$. So in
calculating the $z_i$ the factors of $h$ should cancel. This was not done leading to an 
$h$ dependence through the integrals $\int dz \int dz_p \propto h^2$ and $a(r)=(1+z)\propto (1+h)$ 
i.e. the signal covariance $\propto h^6$. So that the overall spurious $h$ dependence of the signal 
covariance was a factor of $h^{10}$ overall. 

These \emph{errata} resulted in the constraints on $h$ being too 
optimistic in these papers through this strong $h$ dependence. 
Fortunately, for the fiducial surveys considered in these papers, 
the $h$ constraint is dominated by the CMB Planck prior, so that there is a small effect on the 
dark energy parameter constraints. For the larger surveys 
considered in this paper the spurious $h$ 
dependence would lead to the lensing constraint becoming better than the CMB Planck 
prior and as such 
would effect the dark energy constraints by lifting parameter degeneracies. 

\subsection*{3D cosmic shear photometric redshift systematics}
\label{3D cosmic shear photometric redshift systematic}
3D cosmic shear uses the 3D spherical harmonic coefficients characterised by azimuthal $\ell$ and 
radial $k$-modes ($k$ is in $h$Mpc$^{-1}$) given by 
\be
\hat\gamma_{\alpha}(k,\ell)=
\sum_g \gamma^g_{\alpha}kj_{\ell}(k r^g){\rm e}^{-i\ell.\theta^g}.
\ee
for a galaxy at position ($r^g$, $\theta^g$) with shear $\gamma^g$, 
$j_\ell(z)$ are spherical Bessel functions. Note that this results in four independent data vectors 
$\hat\gamma^R_{1}(k,\ell)$, $\hat\gamma^I_{1}(k,\ell)$, $\hat\gamma^R_{2}(k,\ell)$, 
$\hat\gamma^I_{2}(k,\ell)$ where $R$ denotes a real vector and $I$ the imaginary part.

Since the mean of such coefficients is zero it is the covariance of the coefficients that is used to  
extract cosmological information. 
This covariance is a sum of signal and noise terms $C=S+N$ where the signal is given by 
(Heavens et al., 2006; equations 23 to 26) 
\begin{equation}
\label{SPEeq21}
S=\langle\gamma(k,\bell)\gamma^*(k',\bell')\rangle_S =
Q_\ell\,(k,k')\,\delta^D(\bell-\bell')
\end{equation}
where $Q_\ell\,(k,k')$ can be written as
\begin{equation}
\label{SPEeq22}
Q_{\ell}(k,k') = {9 \Omega_m^2 H^4_0 |X_\bell|^2 \over 4\pi^2 c^4}
\int {d\tilde 
k\over \tilde k^2}\, G_\ell(k,\tilde k) G_\ell(k',\tilde k)
\end{equation}
where
\begin{equation}
\label{SPEeq23}
G_\ell(k,\tilde k) \equiv k\int dz\,dz_p\, \bar n_z(z_p) W(z_p)
p(z_p|z) U_\ell(r,\tilde k) j_\ell(kr^0)
\end{equation}
where $p(z_p|z)$ is the photometric redshift probability distribution and
\begin{equation}
\label{SPEeq24}
U_\ell(r,k) \equiv \int_0^r d\tilde r \,{F_K(r,\tilde r)\over
a(\tilde r)} \sqrt{P_\delta(k; \tilde r)} \, j_\ell(k \tilde r)
\end{equation}
where $r=r(z)$ and
\ba
F_K(r,r')&\equiv& \left\{S_k(r-r')/
\left[S_k(r)S_k(r')\right]\right\}\nn
X_\bell &\equiv&{(\bell_y^2-\bell_x^2)+2i\bell_x\bell_y\over \bell^2}.
\ea
$\Omega_m$ is the present dimensionless matter density, $H_0$ is the present Hubble parameter, 
$P_\delta(k; \tilde r)$ is the matter power spectrum, 
$n(z)$ is the source number density distribution and $W(z)$ is an arbitrary weighting function which 
we set to $W(z)=1$.

The photometric redshift probability distribution 
enters the covariance in equation (\ref{SPEeq23}). It acts to damp the signal radially at scales 
approximately equal to and less than the photometric redshift error i.e. at $k$ values of 
$k\geq 2\pi h/(3000\sigma_z)$. Adding extra systematic parameters to $p(z_p|z)$ adds extra parameters 
to the signal of the 3D cosmic shear method but no extra covariance.

\subsection*{3D cosmic shear intrinsic alignments}
\label{3D cosmic shear intrinsic alignments}
Intrinsic alignment effects are included in the 3D cosmic shear method by including additional 
signal covariance terms $C=(S+GI+II)+N$.

The effect of the GI intrinsic alignment effect for 3D cosmic shear is to add a 
further covariance between shear 
and intrinsic ellipticity $\langle \gamma (k,\bell) e^*(k',\bell')\rangle$. 
Using the notation of Heavens et al. (2006), including a convolution over photometric redshift, 
this can be written as
\ba
\label{e5}
GI&=&\langle \gamma (k,\bell) e^*(k',\bell')\rangle\nn
&=&\left(\frac{1}{8\pi^3}\right)\int dzdz_pd^2\btheta
\int dz'dz'_pd^2\btheta'\nn 
&&p(z|z_p)p(z'|z'_p)\bar n(z) \bar n(z') k k' \nn
&&j_{\ell}(kr[z]) j_{\ell'}(k'r[z']){\rm e}^{-i\bell.\btheta}
{\rm e}^{+i\bell'.\btheta'}\langle \gamma (r) e^*(r')\rangle.
\ea
Substituting the parameterised form from equation (\ref{e2}) this can be written as
\ba
\label{e6}
\langle \gamma (k,\bell) e^*(k',\bell')\rangle&=&\left(\frac{A_{\rm GI}}{8\pi^3}\right)
\int dzdz'\nn
&&E(z,z')H_{\ell}(z,k)H_{\ell}(z',k')I(\Omega)
\ea
where $E(z,z')=D(z')D(z-z')/D(z)$ is the lensing efficiency function and 
\ba
\label{e7}
H_{\ell}(z,k)&=&\int dz_p p(z|z_p)\bar n(z)j_{\ell}(kr[z])k\\
I(\Omega)&=&\int_{-\frac{\Delta\theta}{2}}^{+\frac{\Delta\theta}{2}} 
d^2\btheta d^2\btheta'{\rm e}^{-i(\bell.\btheta-\bell'.\btheta')}
\frac{1}{|\btheta-\btheta'|+\theta_0}.
\ea
where $\Omega=\Delta\theta\times\Delta\theta$ is the solid angle of the survey.
The integrals in $I(\Omega)$ can be solved numerically. We use values of  
$A_{\rm GI}=-1.26\times 10^{-7}$ $h {\rm Mpc^{-1}}$ arcmin and $\theta_0=0.90$ arcmin 
taken from Heymans et al. (2006). 

The II intrinsic effect is included by using the following covariance, 
including a convolution over photometric redshift
\ba
II&=&\langle e(k,\bell) e^*(k',\bell')\rangle\nn
  &=&\left(\frac{A_{\rm II}\Omega}{32\pi^5}\right)\int dz dz' M(k,z)M(k',z')\nn
  &&\frac{1}{1+\left(\frac{r[z]-r[z']}{B_{\rm II}}\right)^2}
\ea
where $\Omega$ is the solid angle of the survey, $r(z)$ are comoving distance and
\be
M(k,z)=\int dz_p p(z|z_p)\bar n(z) j_{\ell}(kr)k.
\ee
We use values of  $A_{\rm II}=0.45\times 10^{-3}$ (for mixed galaxies) and $B_{\rm II}=1 h^{-1}$Mpc 
taken from Heymans et al. (2006)

The intrinsic alignments systematics add further covariances to the 3D cosmic shear 
signal and the extra parameters $A_{\rm II}$, $A_{\rm GI}$ and $\theta_0$.

\subsection*{3D cosmic shear distortion}
\label{3D cosmic shear image distortion}
Transformed shear components ($\gamma_1'$,$\gamma_2'$) can be written 
using the parameterisation of equation (\ref{sdist}) as
\ba
\gamma_1'&=&A_{\gamma}[\gamma_1\cos(2\phi)+\gamma_2\sin(2\phi)]+\gamma_{1{\rm bias}}\nn 
\gamma_2'&=&A_{\gamma}[-\gamma_1\sin(2\phi)+\gamma_2\cos(2\phi)]+\gamma_{2{\rm bias}}
\ea 
where $\gamma_{\rm bias}=\gamma_{1{\rm bias}}+i\gamma_{2{\rm bias}}$.
So that, for example, the new $\gamma_2$ estimator 
$\hat\gamma_2'(k,\bell)$ can be written, using the notation of Heavens et al. (2006),
\ba
\label{e12}
\hat\gamma_2'(k,\bell)&=&\left(\frac{2}{\pi}\right)^{1/2}\sum_g \nn
&&k\{A_{\gamma}[-\gamma_1\sin(2\phi)+\gamma_2\cos(2\phi)]+\gamma_{2{\rm bias}}\}\nn
&&j_{\ell}[kr(z)]{\rm e}^{-i\bell.\btheta}.
\ea
Taking the covariance results in four terms respectively proportional to
\ba
\label{A2}
&&A^2_{\gamma}\langle\gamma_1\gamma^*_1\rangle\sin^2(2\phi)\nn
&&A^2_{\gamma}\langle\gamma_2\gamma^*_2\rangle\cos^2(2\phi)\nn
&&A_{\gamma}\langle\gamma_{2{\rm bias}}\gamma^*_2\rangle\cos(2\phi)\nn
&&A_{\gamma}\langle\gamma_2\gamma^*_{2{\rm bias}}\rangle\cos(2\phi)
\ea 
to first order in $\gamma_{2{\rm bias}}$.

In the calculation, in
order in save computational time, the $\gamma^R_2(k,\bell)$ component is
chosen as representative so that in the signal covariance, equation (\ref{SPEeq22}), 
$X_{\bell}=2|\bell|^2\cos(\phi_{\ell})\sin(\phi_{\ell})$ where
$\phi_{\ell}$ is the angle of the $\bell$ vector in the ($\ell_x$,
$\ell_y$) phase space. We only consider $\ell_x\geq 0$ to avoid double
counting. By choosing $\phi_{\ell}=\pi/4$ the prefactor becomes
$X_{\bell}=|\bell|^2$, the Fisher matrix is then integrated over all
modes in a given shell using the $\gamma^R_2(k,\bell)$ component as
representative 
\be
\F_{\alpha\beta}=g\int_{-\pi/2}^{\pi/2}d\phi_{\ell}\int d\ell \ell \F_{\alpha\beta}(\ell)
\ee
where  $\F_{\alpha\beta}(\ell)={1\over 2}{\rm
  Tr}\left[(C^\bell)^{-1}C^\bell_{,\alpha}(C^\bell)^{-1}C^\bell_{,\beta}\right]$. 
The density of states in $\ell$-space due   
to the survey size is $g=\Delta\Omega/(2\pi)^2$ so that 
\be
\F_{\alpha\beta}=\frac{\Delta\Omega}{4\pi}\int d\ell \ell
\F_{\alpha\beta}(\ell). 
\ee
To test this approximation a full Fisher matrix calculation was done
over all $\phi_{\ell}$ where $\F_{\alpha\beta}(\ell)\rightarrow
\F_{\alpha\beta}(\ell,\phi_{\ell})$, for all four data vectors, the
different $\phi_{\ell}$ 
dependence comes from the $X_{\bell}$ prefactor. The dark energy
parameter errors were found be in agreement to within $\pm 0.001$ 
and since the computational time is $4\times N_{\phi}$ larger for the full
calculation (where $N_{\phi}$ is the total number of modes calculated in
$\ell$-space) the approximation is used for all predictions in
this paper. 

Since in the calculation we use $\phi_{\ell}=\pi/4$ 
as a representative point in the $\bell$ phase space and integrate over $\bell$ the terms in 
equation (\ref{A2}) can 
be reduced since $\langle\gamma_1\gamma^*_1\rangle\rightarrow 0$. Also, since we assume that 
$\gamma_{2{\rm bias}}=$constant the expression reduces further since 
$\langle\gamma_{2{\rm bias}}\gamma^*_2\rangle=\gamma_{2{\rm bias}}\langle\gamma^*_2\rangle=0$.
So the signal is simply multiplied
by an extra factor of $A^2_{\gamma}\cos^2(2\phi)$. In the shot noise calculation 
$\langle\gamma_1\gamma^*_1\rangle=\langle\gamma_2\gamma^*_2\rangle=\sigma^2_e$ since 
$[\langle\gamma_1\gamma^*_1\rangle\sin^2(2\phi)+
\langle\gamma_2\gamma^*_2\rangle\cos^2(2\phi)]=[\cos^2(2\phi)+\sin^2(2\phi)]\sigma^2_e=\sigma^2_e$, 
so the shot noise is simply multiplied by $A^2_{\gamma}$. A caveat is that the assumption that the 
mean signal is zero 
would break down if $\gamma_{\rm bias}$ became too large. Here we assume that the 
assumption that the mean signal is zero is still valid since $\gamma_{\rm bias}\ll \gamma$ which will 
be valid to first order in $\gamma_{\rm bias}$. 

Since the extra shear systematic parameters are in the signal, the calibration $A_{\gamma}$ and the 
angle $\phi$ can be marginalised over and no extra covariance terms are added to the method.

\end{document}